\documentclass[prb,twocolumn,preprintnumbers,amsmath,amssymb]{revtex4}

\usepackage{graphicx}

\begin{document}
\title{ \bf    Elementary excitations, Spectral weights and Experimental signatures in a Supersolid and
a Fulde - Ferrell-Larkin-Ovchinnikov  state }
\author{ \bf  Jinwu Ye  }
\affiliation{ Department of Physics, The Pennsylvania State University, University Park, PA, 16802       }
\date{\today}
\begin{abstract}
   We construct a Ginsburg Landau (GL ) theory to study the phases of liquid, solid, superfluid, especially  a possible supersolid  and phase transitions
   among these phases in a unified framework. In this GL, we put the two competing orders between the solid component and the superfluid component
   on the same footing. We only introduce two parameters: $ v $ which is the repulsive interaction between the normal component and the
   local superfluid mode and  $ g $ which is a periodically changing chemical potential for the local superfluid mode.
   The microscopic origins of $ v $ and $ g $ are given.
   By using this GL action, we study the superfluid to supersolid transition,
   normal solid to the supersolid transition and analyze the
   conditions for the existence of the supersolid.
  The non-classical rotational inertial (NCRI) in the SS state
  is calculated and found to be isotropic in $ bcc $ and $ fcc $
  lattice, but weakly anisotropic in $ hcp $ lattice.
  We study the elementary low energy excitation inside a
  supersolid. We find that there are one upper  branch and one lower branch longitudinal "supersolidon"
  modes inside the SS,  while the transverse modes in the SS stay the same as those inside the NS.
  We also determine the corresponding spectral weights  of the two branches.
  We work out the experimental signatures of these elementary excitations in
  Debye-Waller factor, density-density correlation, vortex loop
  interaction and specific heat. The estimated excess entropy due
  to vacancies seems consistent with data measured in the specific  experiment in Helium 4.
  Detecting the two supersolidon modes by various
  equilibrium and thermodynamic experiments such as X-ray scattering, neutron scattering,
  acoustic wave attenuation and heat capacity can prove or
  disapprove the existence of a supersolid in Helium 4.   A toy model of supersolid wavefunction is analyzed.
  The  difference and similarities with lattice supersolid  are clearly demonstrated.
  Elementary excitations inside a Fulde - Ferrell-Larkin-Ovchinnikov (FFLO) state of superfluid
  are also discussed.

\end{abstract}

\maketitle

\section{Introduction}

     A solid can not flow. It breaks a continuous translational
     symmetry into a discrete lattice translational symmetry.
     There are low energy lattice phonon excitations in the solid.
     While a superfluid can flow even through narrowest channels
     without any resistance. It breaks a global $ U(1) $ phase
     rotational symmetry and has the off-diagonal long range order (ODLRO)
     \cite{yang}. There are low energy superfluid phonon excitations
     in the superfluid. A supersolid is a state which breaks both
     the continuous translational symmetry and the global $ U(1) $
     symmetry, therefore has both the crystalline  order and the
     ODLRO.  The possibility of a supersolid phase in $^{4}He$ was
     theoretically speculated in 1970 \cite{and,ches,leg,sas}. Andreev  and Lifshitz
     proposed the Bose-Einstein condensation of vacancies as the mechanism of
     the formation of supersolid \cite{and}. Chester wrote down a
     wavefunction which has both ODLRO and crystalline order and
     also speculated that a supersolid cannot exist without
     vacancies or interstitial \cite{ches}. Leggett proposed that solid
     $^{4}He$ might display Non-Classical Rotational Inertial (NCRI)
     which is  a low temperature reduction in the rotational moment of inertia
     due to the superfluid component of solid $^{4}He$ \cite{leg,sas,gas}.
     Leggett also suggested that  quantum tunneling of He atoms
     between neighboring sites in a crystal can also lead to a
     supersolid even in the absence of vacancies. Over the last 35
     years, a number of experiments have been designed to search for
     the supersolid state without success. However, recently, by
     using torsional oscillator measurement, a PSU group lead by Chan observed a marked $ 1 \sim 2 \% $  NCRI of solid $^{4}He$
     at $ \sim 0.2 K $, both when
     embedded in Vycor glass and in bulk $^{4}He$ \cite{chan}.
     The authors suggested that the supersolid state of $^{4}He$ maybe responsible
     for the NCRI.  The
     PSU group also detected about $ 10^{-4} $ reduction in the
     rotational inertial in solid $ H_{2} $ \cite{h2}, but by performing blocked cell experiments, the authors concluded
     that the reduction is {\em classical} due to the motion of $ HD $ impurities
     clustering. So the PSU group did not find any NCRI in solid hydrogen \cite{h2}.

    The PSU experiments inspired  extensive
    both theoretical \cite{micro,macro,dor,ander} and experimental \cite{annealing,massflow,melt} interests
    in the very intriguing  supersolid phase of $^{4}He$. Now it was widely believed that disorders
    played key roles in the PSU experiments.
    There are two kinds of complementary theoretical approaches.
    The first is the microscopic
    numerical simulation \cite{micro}. The second is the phenomenological
    approach \cite{macro,dor,ander,qglprl}.  In this paper, we assume no disorders in our GL theory and study all the possible
    phases in different parameter regimes of the GL model.
    A solid has the order in the number density, while a superfluid has the
    order in the phase. The two phases are in totally different
    extremes of the state of matter. How to combine the two opposite
    extremes into a new state of matter, supersolid, is an
    important and interesting topic on its own.
    It is widely believed that the
    only chance to get a supersolid is that the
    solid is not {\em perfect}, namely, it is an incommensurate solid which has defects such as
    vacancies or interstitials, so the total number of bosons can fluctuate to give
    some rooms for the invasion of superfluid. Due to large zero point motions, there are
    indeed rapid exchanging between local $ ^{4}He $ atoms in bulk $ ^{4}He $,
    but this local process will not lead to a global phase ordering.
    It was estimated from  both X-ray scattering experiments \cite{simmons} and some
    microscopic calculations \cite{micro} that the thermal excitation
    energies of a vacancy and interstitial are $ \epsilon_{v} \sim 10 \ K, \epsilon_{i} \sim 40  \ K  $,
    so thermal fluctuations favor vacancies over interstitials.
    Because both have very high energy, so the thermal generated vacancies and interstitials are irrelevant
    around $ 45 \sim 200 \ mK $. However, it is still possible that there
    are quantum fluctuation generated vacancies and interstitials
    even at $ T=0 $. It is still not known if quantum fluctuations favor
    vacancies over interstitials or not\cite{ander}.

    In this paper, we will construct a phenomenological Ginsburg Landau ( GL ) theory to
    study all the possible phases and phase transitions in helium 4
    system. We identify order parameters, symmetry and symmetry breaking
    patterns in all the phases. Particularly, we will address the
    following two questions: (1) What is the condition for the existence
    of the SS state ? (2) If the SS exists, what are the properties of the
    supersolid to be tested by possible new experiments.
    Let's start by reviewing all the known phases
    in $^{4}He$. The density of a normal solid (NS) is defined as $ n( \vec{x}
     )=  n_{0}  + \sum^{\prime}_{\vec{G}} n_{\vec{G}} e^{i \vec{G} \cdot
     \vec{x} } =n_{0}+ \delta n( \vec{x} )  $ where $ n^{*}_{\vec{G}}=n_{- \vec{G}}$ and $ \vec{G} $ is any non-zero reciprocal lattice vector.
     In a normal liquid (NL), if the static liquid structure factor $ S(k) $ has its first maximum peak
     at $ \vec{k}_{n} $, then near $ k_{n} $, $ S(k) \sim  \frac{1}{r_{n} + c ( k^{2}-k^{2}_{n} )^{2} } $.
     If the liquid-solid transition is weakly first order, it is known that
     the classical free energy to describe  the NL-NS transition is \cite{tom}:
\begin{eqnarray}
  f_{n} & = &  \sum_{\vec{G}} \frac{1}{2}  r_{\vec{G}} | n_{\vec{G}} |^{2}
    -w  \sum_{\vec{G}_1,\vec{G}_2,\vec{G}_3} n_{\vec{G}_1}
    n_{\vec{G}_2} n_{\vec{G}_3} \delta_{ \vec{G}_1 + \vec{G}_2 +
    \vec{G}_3,0 }  \nonumber   \\
    & + &  u_{n} \sum_{\vec{G}_1,\vec{G}_2,\vec{G}_3,\vec{G}_4 } n_{\vec{G}_1}
    n_{\vec{G}_2} n_{\vec{G}_3} n_{ \vec{G}_4 }  \delta_{ \vec{G}_1 + \vec{G}_2 +
    \vec{G}_3+ \vec{G}_4, 0 } + \cdots
\label{sl}
\end{eqnarray}
    where $ r_{\vec{G}}=r_{n} + c ( G^{2}-k^{2}_{0} )^{2} $ is the
    tuning parameter controlled by the pressure or temperature.
    Note that the average density $ n_{0} $ does not enter in the free energy.
    Because the instability happens
    at finite wavevector, Eqn.\ref{sl} is an expansion in terms of small parameter $ n_{\vec{G}} $ alone,
    it is {\em not a gradient expansion }!
    The GL parameters $
    w $ and $ u_{n} $ may be determined by fitting the theoretical predictions with
    experimental data.
    It is easy to see that Eqn.\ref{sl} is
    invariant under $ \vec{x} \rightarrow \vec{x} + \vec{a}, n(
    \vec{x} ) \rightarrow n( \vec{x} +\vec{a} ), n ( \vec{G} )
    \rightarrow n ( \vec{G} ) e^{ i \vec{G} \cdot \vec{a} } $ where
    $ \vec{a} $ is any vector. In the NL, $ \langle n ( \vec{G} )\rangle =0 $, the translational symmetry is respected.
    In the NS, $ \langle n ( \vec{G} ) \rangle \neq 0 $, the symmetry is broken down to translational
    invariance under only a lattice constant $ \vec{a}= \vec{R}, \vec{G} \cdot \vec{R}= 2 \pi n,
    n( \vec{x} ) \rightarrow n( \vec{x} +\vec{R} ), n ( \vec{G} )
    \rightarrow n ( \vec{G} ) $.
    As shown in the Fig.1, the NL to NS transition only happens at finite
    temperature, so the classical theory is valid.
    Note that due to {\em the lack} of the $ Z_{2} $ symmetry
    of $ \delta n(\vec{x} ) \rightarrow -\delta n(\vec{x} ) $, namely,
    $ n_{\vec{G}} \rightarrow - n_{\vec{G}} $, there is always a
    cubic $ w $ term which makes the the NL to NS a 1st order transition.
    The $ u_{n} $ term which is invariant under the $ Z_{2} $ symmetry is needed for the stability reason.

    Of course, the Superfluid to Normal Liquid transition  at finite temperature in the Fig.1 is the 3d
    XY transition described by \cite{3dxy}:
\begin{equation}
  f_{\psi}  =  K | \nabla \psi |^{2} + t | \psi |^{2} +  u  |\psi |^{4}  + \cdots
\label{sfnormal}
\end{equation}
   where $ \psi $ is the complex order parameter and $ t $ is the
   tuning parameter controlled by the temperature or pressure.  Eqn.\ref{sfnormal} is invariant under the global $ U(1) $
   symmetry $ \psi \rightarrow \psi e^{ i \theta } $. In the NL, $ \langle \psi \rangle =0 $, the
   symmetry is respected. In the SF,  $ \langle \psi \rangle \neq 0 $, the symmetry
   is broken. The GL parameters $
    K, t, u $ may be determined by fitting the theoretical predictions with
    experimental data. In this paper, we always assume $ u > 0 $.

    The coupling between $ n ( \vec{x} ) $ and $ \psi( \vec{x} ) $ consistent
    with all the  symmetry  can be written down as:
\begin{equation}
  f_{int} =   g \delta n ( \vec{x} ) | \psi( \vec{x} ) |^{2}
             + v ( \delta n( \vec{x} ))^{2} | \psi( \vec{x} ) |^{2} + \cdots
\label{int}
\end{equation}
   where  $ \delta n ( \vec{x} ) = n ( \vec{x} )-n_{0}
   = \sum^{\prime}_{\vec{G}} n_{\vec{G}} e^{i \vec{G} \cdot \vec{x} }
   $. Note that the average density $ n_{0} $ does not enter in the interaction either.
   Eqn.\ref{int} is an expansion in terms of two small parameters $ \delta n ( \vec{x}
   ) $ and $ \psi( \vec{x} ) $.
   The $ \cdots $ include the other terms in higher odd and even powers of $ \delta n( \vec{x} ) $ which are
   sub-leading to the $ g $ and $ v $ term. In an effective GL theory, $ n( \vec{x} ) $ and $ \psi (\vec{x} ) $
     emerge as two independent order parameters.

      Eqn.\ref{int} is invariant under both the translational symmetry $ \vec{x} \rightarrow \vec{x} +
   \vec{a},  n( \vec{x} ) \rightarrow n( \vec{x} +\vec{a} ),  \psi ( \vec{x} ) \rightarrow
   \psi( \vec{x} +\vec{a} ) $ and the global $ U(1) $ symmetry $ \psi \rightarrow \psi e^{ i \theta }
   $. Note that it is important to keep both $ g $ and $ v $ term in
   the Eqn.\ref{int}, because the $ g $ term changes the sign, while the $
   v $ term is invariant under the Particle-Hole ( PH ) transformation
   $ \delta n(\vec{x}) \rightarrow -  \delta n(\vec{x}) $,
   so the sign of $ g $ makes a difference ! Due to the two competing orders between the density fluctuation represented
   by $ ( \delta n (\vec{x}) )^{2} $ and the local superfluid mode
   represented by $ |\psi|^{2} $, we expect $ v $ to be always positive and is an increasing function of the pressure $ p $. The positive $
   v $ term is also needed for the stability reason.
    On the other
   hand, we can view $ g $ as a {\em periodic } chemical potential {\em with average zero} acting on $ \psi $.
   It is easy to see the coupling is attractive $ g_{v} < 0 $ for vacancies,
   but repulsive $ g_{i} > 0 $ for interstitials. From
   Eqn.\ref{int}, we can classify three kinds of solids:
   If $ g=0 $, the C-NS has the P-H symmetry, let's call this kind of PH
   symmetric C-NS as NS-PH. In general, $ g \neq 0 $,
   so there is no particle-hole symmetry in the C-NS,
   there are still two kinds: ( 1) vacancy like C-NS where the excitation energy of a
   vacancy is lower than that of an interstitial, named NS-v. (2)
   interstitial like C-NS where the excitation energy of an
   interstitial is lower than that of a vacancy, named NS-i.
   We expect that in contrast to $ v $, $ g $ is an intrinsic parameter of  solid
   Helium 4 which depends on the mass of  a $ ^{4} He $ atom and the  potential between the $ ^{4} He $
   atoms, but not sensitive to the pressure $ p $. In SS-v and SS-i, $ \psi (\vec{x} )
   $ stands for vacancies and interstitials respectively.
   The total density of the system is $ n_{t}(x)= n(x) \pm
   |\psi(x)|^{2} $  for SS-i and SS-v respectively where $ n(x) $ is the density of number of solid sites and $  |\psi(x)|^{2} $ is the superfluid density.
   Both were treated as independent order parameters in the GL theory.

    The GL equations \ref{sl},\ref{sfnormal},\ref{int} are invariant under both the translational
     symmetry $ \vec{x} \rightarrow \vec{x} +
     \vec{a},  n( \vec{x} ) \rightarrow n( \vec{x} +\vec{a} ), n ( \vec{G} )
    \rightarrow n ( \vec{G} ) e^{ i \vec{G} \cdot \vec{a} },  \psi ( \vec{x} ) \rightarrow
     \psi( \vec{x} +\vec{a} ) $ and
     the global $ U(1) $ symmetry $ \psi \rightarrow \psi e^{ i \theta }
     $. In a NL, $ n_{ \vec{G} } = 0, \langle \psi\rangle =0 $. In a SF, $ n_{ \vec{G} } =  0, \langle \psi\rangle \neq 0
     $. In a NS, $ n_{ \vec{G} } \neq 0,
     \langle \psi\rangle =0 $. While in a  supersolid, $  n_{ \vec{G} } \neq 0,
     \langle \psi\rangle \neq 0  $. From the normal liquid ( NL ) side, one can approach both the solid
      and the superfluid.
      Inside the NL,  $ t > 0 $, $ \psi $ has a gap, so can be integrated
      out from Eqn.\ref{int}, we recover the solid-liquid transition tuned by $ r_{\vec{G}} $ in Eqn.\ref{sl} ( Fig.2 ).
      Inside the NL $ \langle n( \vec{x} )\rangle =n_{0} $, the density fluctuations of $ \delta n(x) $ is massive,
      so can be integrated out from Eqn.\ref{int},
      then we recover the NL to SF transition tuned by $ t $ in Eqn.\ref{sfnormal} ( Fig.1 ).

    Quantum fluctuations can be incorporated  by (1)  $ n(
    \vec{x} ) \rightarrow n( \vec{x},\tau) $ and including $
    \frac{1}{2} \rho_{n} ( \partial_{\tau} n )^{2} $ in
    Eqn.\ref{sl}. (2) $ \psi(
    \vec{x} ) \rightarrow \psi ( \vec{x},\tau) $ and including $
    \psi^{\dagger}( \vec{x},\tau)  \partial_{\tau} \psi( \vec{x},\tau) $ in Eqn.\ref{sfnormal}.
    (3) Due to the lack of particle-hole symmetry in the normal
    solid, including additional terms like $ \delta n( \vec{x}, \tau )
    \psi^{\dagger}(\vec{x}, \tau ) \partial_{\tau} \psi(\vec{x}, \tau ) $ in  Eqn. \ref{int}.
    We will explicitly consider the quantum fluctuations in section III where we will discuss the quantum phase transition from
    the superfluid to the C-NS ( either NS-v or NS-i ).
   However,  as shown in this paper, the quantum fluctuation terms are very
   important in the zero temperature quantum phase transition from SS-v ( SS-i ) to
   NS-v ( NS-i ) driven by the pressure ( see Fig.2 ) and
   to determine the low energy excitation spectra in a given phase.



    The rest of the paper is organized as the following:   In sec. II, for
    the first time, we construct a quantum Ginsburg-Landau action to study SF to the NS
    transition and also explicitly establish the Feynman relation
    from the QGL. In Sec.III, taking into
    account the couplings between the $ n $ and $ \psi $ sector
    encoded in Eqn.\ref{int}, we study the superfluid to
    supersolid transition  which is a simultaneous combination of the SFDW transition in $ \psi $ sector
    driven by the roton condensation  at $ k_{0}=k_{r} $ and
    the NS transition in the $ n $ sector driven by the divergence of the structure function
    at $ k_{0}=k_{r}=k_{n} $  discussed in Sec.II.
    We also sketch the global
    phase diagrams to be confirmed and analyzed in the following
    sections.
    In Sec.IV, we approach the SS phase  from the
    NS side and discuss the SFDW in SS-v and SS-i respectively.
    We analyze carefully the conditions for the existence of SS-v.
    We explicitly show that the SS-v is the possible ground state
    when the $ g_{v} $ is sufficiently negative ( Fig.3).
    The SS-i is less likely, but we still analyze SS-i in the same
    footing as SS-v. We also classify several common SS-i lattice structures.
   In Sec. V, by renormalization group analysis, we study the
   universality class of zero temperature quantum phase transition
   from normal solid (NS ) to supersolid (SS) driven by the
   pressure. In Sec. VI, we calculate the NCRI's
    in both the SF and SS states and  find the NCRI in the $ hcp $ SS lattice may be weakly anisotropic.
   In Sec.VII, we  work  out the non-topological elementary
   excitations  and corresponding spectral weights inside the supersolids in both the isotropic solid case
   and the hcp lattice structure case. Then in Sec.VIII, we study
   topological elementary excitations, namely, vortex loops  and vortex lines in the SS
   by performing a duality transformation to the vortex loop representation.
   We also estimate the very large vortex core size and low critical velocity in the SS state.
   Then in the following sections, we study the experimental signatures of these low
   energy elementary excitations:
   In Sec. IX, we  make key predictions on the elastic X-ray scattering
   amplitudes from all SS-v  and SS-i structures classified in Sec. IV by
   calculating the Debye-Waller factor and the density-density
   correlation function.  In Sec. X, we study the specific heat in the SS.
   In Sec. XI,  inspired by the results achieved in Sec. VII, we discuss the elementary excitations inside a
   Fulde - Ferrell-Larkin-Ovchinnikov (FFLO) state of superconductors.
    Conclusions are summarized in the final section XII.
    In the appendix A, we discuss the properties of a tight-binding {\em toy} SS-v wavefunction.
    In the appendix B, we compare the properties of the SS with
    those of SS on extended boson Hubbard model in a lattice.
    Two short reports of these results appeared in \cite{qglprl,epl}.

\section{ Superfluid to normal solid transition }

     Let us start with the SF to the NS transition.
     In the superfluid state, if the multi-quasiparticle part can be
     neglected in the dynamic structure factor,
     the Feynman relation between the Landau quasi-particle dispersion relation in the $ \psi $ sector
     and the equal time structure factor in the $ n $ sector holds:
\begin{equation}
    \omega( q ) = \frac{q^{2}}{ 2m S(q) }
\label{fey}
\end{equation}

      In the $ q \rightarrow 0 $ limit, $ S(q) \sim q $, $ \omega(q)
      \sim q $  recovers the superfluid phonon spectrum near $ q=0 $ in Fig.1b.
      The first maximum peak in $ S(q) $ corresponds to the roton minimum in $
      \omega(q) $ in the $ \psi_{2} $ sector ( Fig.1b ), namely, $ k_{n}=k_{r} $.
      As one increases the pressure $ p $, the interaction
      $ u $ in Eqn.\ref{int} also gets bigger and bigger, the first maximum peak of $ S(q) $ increases,
      the roton minimum $ \Delta_{r} $ gets smaller and
      smaller\cite{roton2}. Across the critical pressure $ p=p_{c} $, there are
      the two possibilities. (1) The resulting solid is a commensurate solid, then $ \langle \psi \rangle=0 $.
      In the NS, there is no remanent of the roton inside the SF,
      the supersolid phase does not exist as an equilibrium
      ground state. This is the SF to the C-NS transition.
      As to be shown in section IV, this happens when $ | g | $ is sufficiently small ( Fig.1 ). (2)
      The resulting solid at high pressure is an incommensurate solid
      with zero point quantum fluctuations generated vacancies or interstitials whose condensation leads to the
      formation of the SS-v and SS-i respectively.
      In section IV, this happens when $ | g | $ is sufficiently large ( Fig.2 ).
      We will discuss case (1) in this section, then the most interesting case
      (2) in the next section.

    The effective action inside the SF is :
\begin{equation}
 {\cal L}[ \delta n, \theta ]= i \delta n  \partial_{\tau} \theta +
          \frac{ 1 }{2}  \rho_{s} ( \nabla \theta )^{2}
          + \frac{1}{2} \delta n V_{n} (\vec{q} ) \delta  n
\label{nnxy}
\end{equation}
   where $ \rho_{s} $ is the superfluid density and
   $ V_{n}(q)= a- bq^{2}+ cq^{4} $ with $ a>0, b>0 $ is the density-density
   interaction between the $^{4} He $ atoms.

    In the SF state, it is convenient to integrate out
    $ \delta n $ in favor of the phase field $ \theta $ to get  the phase representation
\begin{equation}
    {\cal L}[ \theta ] = \frac{1}{ 2 V_{n}( \vec{q} ) } ( \partial_{\tau} \theta )^{2} +
     \frac{ 1 }{2}  \rho_{s} ( \nabla \theta )^{2}
\label{xy}
\end{equation}
     where one can identify the compressibility $ \kappa^{-1} = \lim V_{n}(q) |_{q \rightarrow 0 }=a $.
     The dispersion relation of the superfluid modes including higher
     orders of momentum can be extracted \cite{blqh}:
\begin{equation}
     \omega^{2} = [ \rho_{s}  V_{n}( \vec{q} ) ] q^{2}= 2 \rho_{s} q^{2} (  a- bq^{2}+ cq^{4} )
\label{dip}
\end{equation}
     where one can see the superfluid phonon velocity $ v^{2} = \rho_{s}/\kappa $.

     It is easy to see that the dispersion relation indeed has the
     form shown in Fig.1b with a roton minimum.
     Because the original instability
     comes from the density-density interaction $ V_{n}( \vec{q} ) $ ,
     it is convenient to integrate out the phase field in favor of the
     density fluctuation operator $ \delta n $.
     Neglecting the vortex excitations in $ \theta $ and integrating
     out the $ \theta $ in Eqn.\ref{nnxy} leads to:
\begin{equation}
 {\cal L}[ \delta n ]=
    \frac{1}{2} \delta n (-\vec{q},-\omega_{n} ) [ \frac{ \omega^{2}_{n} }{ \rho_{s} q^{2}}
    + V_{n} (\vec{q} ) ] \delta n (\vec{q},\omega_{n}  )
\label{density}
\end{equation}
    where we can identify the dynamic pseudo-spin density-density correlation
    function $ S_{n}(\vec{q},\omega_{n} )= \langle \delta n (-\vec{q},-\omega_{n} ) \delta n
    (\vec{q},\omega_{n} ) \rangle =  \frac{  \rho_{s} q^{2} }{ \omega^{2}_{n} +
   v^{2}(q) q^{2} } $
    where $ v^{2}(q)=  \rho_{s} V_{n} (q) $ is the " momentum dependent " spin wave velocity.
    Obviously $ v^{2}(q=0)= v^{2} = \rho_{s}/\kappa $.


    From the pole of the dynamic density-density correlation
    function, we can identify the speed of sound
    wave which is exactly the same as the spin wave velocity.
    This should be expected.
    From the analytical continuation $ i \omega_{n} \rightarrow \omega+ i\delta $ and taking the imaginary part, we can identify the dynamic
    structure factor: $ S^{>}_{n}( \vec{q},\omega ) = S_{n}(q) \delta ( \omega
    -v(q)q ) $ where $ S_{n}(q)= \rho_{s} q \pi/2v(q) $ is the equal time
    density correlation function shown in Fig.1b. As $ q  \rightarrow 0, S_{n}(q) \rightarrow q $.
    The sum rules are:
 \begin{eqnarray}
      \int^{\infty}_{0} d \omega  S^{>}_{n}( \vec{q},\omega ) & = & \rho_{s} q \pi/2v(q) =  S_{n}(q)  \nonumber   \\
      \int^{\infty}_{0} d \omega \omega  S^{>}_{n}( \vec{q},\omega ) & = & \rho_{s} q^{2} \pi/2       \nonumber   \\
      \int^{\infty}_{0} d \omega  S^{>}_{n}( \vec{q},\omega )/\omega & = & \rho_{s}  \pi/2v^{2}(q)
\label{sumrule}
\end{eqnarray}
    where the first one just defines the equal time density correlation function, the second one gives the $ f $ sum rule which can be used to define
    the superfluid density $ \rho_{s} $,
    the third one can be used to define the compressibility $ \kappa $:
    $  \int^{\infty}_{0} d \omega  S^{>}_{n}( \vec{q},\omega )/\omega |_{q \rightarrow 0} = \kappa  \pi/2   $.

    The {\sl Feynman relation } which relates the dispersion relation to the equal time structure factor
    is:
\begin{equation}
     \omega (q) =  \frac{  \int^{\infty}_{0} d \omega  \omega S^{>}_{n}( \vec{q},\omega )}
     {\int^{\infty}_{0} d \omega   S^{>}_{n}( \vec{q},\omega ) }=  \frac{ \rho_{s} \pi  q^{2} }{ 2 S_{n}(q) }
\label{fey2}
\end{equation}
    which takes exactly the same form as Eqn.\ref{fey} if we
    identify $  \rho_{s} \sim 1/m $ with $ m $ the mass of
    $^{4} He $ mass. Therefore, we recovered the Feynman relation
    from our GL theory Eqn.\ref{nnxy} which gives us the confidence
    that Eqn.\ref{nnxy} is the correct starting action to study the the SF
    to SS transition.
    The density representation Eqn.\ref{density} is {\em dual} to the
    phase representation Eqn.\ref{xy}.
    However, the phase representation Eqn.\ref{xy} contains explicitly the superfluid order
    parameter $ \psi \sim e^{i \theta } $ which can be used
    to characterize the superfluid order in the SF phase.
    While in Eqn.\ref{density},   because the order parameter $ \psi $ is integrated out, the superfluid
    order is hidden, the signature of the
    superfluid phonon mode is encoded in the density sound mode, so it is not as effective as the phase representation in describing
    the SF state. However, as shown in the following, when describing the transition
    from the ES to the NS, the density representation Eqn.\ref{density} has a big advantage
    over the phase representation.

     Because the instability is happening at $ q=q_{0} $ instead of at $ q=0 $, the vortex excitations in
     $ \theta $ remain {\em uncritical} through the SF to SS transition. Integrating them out will generate
     interactions among the density $ \delta n $:
\begin{eqnarray}
 {\cal L}[ \delta n ] & = &
    \frac{1}{2} \delta n (-\vec{q},-\omega_{n} ) [ \frac{ \omega^{2}_{n} }{ \rho_{s} q^{2}}
    + V_{n} (\vec{q} ) ] \delta n (\vec{q},\omega_{n} )  \nonumber  \\
    & - &  w ( \delta n )^{3} + u ( \delta n  )^{4} + \cdots
\label{densityint}
\end{eqnarray}
    where the momentum and frequency conservation in the quartic
    term is assumed. Note that the $  ( \omega_{n} /q )^{2} $ term in the first
    term stands for the quantum fluctuations of $ \delta n $ which is absent in the classical
    NL to NS transition Eqn.\ref{sl}. Because of the lack of $ Z_{2} $ exchange symmetry, there is a cubic term in
    Eqn.\ref{densityint}.  Expanding $ V_{n}(q) $ near the roton minimum $ q_{0} $
    leads to the quantum Ginsburg-Landau action to describe the SF to the NS transition:
\begin{eqnarray}
 {\cal L}[ \delta n ] &  = &
    \frac{1}{2} \delta n [ A_{n} \omega^{2}_{n}
    + r+ c( q^{2}-q^{2}_{0} )^{2} ] \delta n   \nonumber  \\
    & -  & w ( \delta n )^{3} + u ( \delta n )^{4} + \cdots
\label{sfdensity}
\end{eqnarray}
    where $ r \sim p_{c1}-p $ and  $  A_{\rho} \sim \frac{ 1 }{ \rho_{s} q_{0}^{2}} $ which
    is non-critical across the transition. Just like Eqn.\ref{sl},
    because the instability happens at the finite wavevector $
    q=q_{0} $, Eqn.\ref{sfdensity} is {\em not a gradient expansion}, but
    an expansion in terms of the small order parameter $ \delta n
    $. Again the average density $ n_{0} $ does not appear in
    Eqn.\ref{sfdensity}.
    The generic transition driven by the collapsing of roton minimum is from SF to NS instead of
    from the SF to the supersolid ( SS ). In the SF, $ r > 0, \langle \psi \rangle \neq 0,  \langle \delta n \rangle =0 $,
    In the NS, $ r < 0, \langle \psi \rangle = 0, \langle \delta n \rangle = \sum^{\prime}_{\vec{G}}  n_{\vec{G}} e^{i \vec{G} \cdot
    \vec{x} } $ where $ \vec{G} $ are the shortest reciprocal lattice vector of the resulting lattice.
The corresponding phase diagram  for the SF to the NS transition is
shown in Fig.1.

\vspace{0.25cm}

\begin{figure}
\includegraphics[width=8cm]{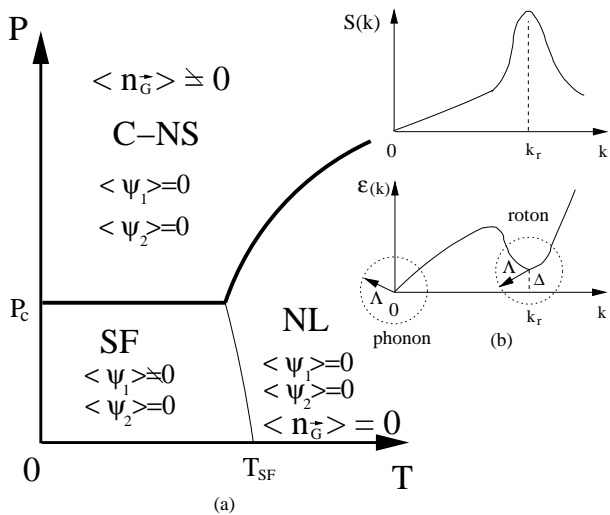}
\caption{  (a)  The theoretical phase
   diagram of GL model Eqns.\ref{sl},\ref{sfnormal} and \ref{int} in $  P  $ versus $ T $ plane.
   This phase diagram only happens when $ |g| $ is sufficiently
   small, so the normal solid is a C-NS ( Fig.3 ). $ T $ controls thermal
   fluctuations, while $ p $ tunes quantum fluctuations.
   SF is the superfluid phase, C-NS is the commensurate normal solid phase which still does not have
   the P-H symmetry, it is likely to be a vacancy-like C-NS ( NS-v ). NL is the normal liquid phase. The supersolid phase in Fig.2 is absent.
   The $ \langle \psi_{1} \rangle $ and $\langle \psi_2  \rangle $ stand for the condensations in zero momentum and the first reciprocal lattice vectors respectively.
   Thick ( thin ) lines are 1st ( 2nd ) order transitions. The critical
   temperatures $ T_{SF} $ of NL to SF transition
   drops slightly as the pressure $ p $ increases because of the quantum fluctuations.
   The SF to the C-NS transition is described by the QGL Eqn.\ref{sfdensity}.
   (b) The structure factor and the separation  of low ( phonon ) and high ( roton ) momenta
   regime in the dispersion relation in the SF. The Feynman
   relation between the two curves are explicitly derived in Eqn.\ref{fey2}. }
\label{fig1}
\end{figure}

\section{ Superfluid to supersolid transition  }

      In the last section, we discuss  the SF to the NS
      transition with the order parameter  $ \delta n $. In this
      section, we discuss the SF to the SS transition with both
      order parameters $ \psi $ and $ \delta n $. Then we have to
      also  include the couplings between $ \delta n $
      and $ \psi $ sector encoded in Eqn.\ref{int}.
     Starting from the SF side with increasing pressure,
     we develop the theory based on the two facts (1) there is a roton minimum in
     the superfluid state (2) The instability to solid formation is
     driven by the gap diminishing at the roton minimum.
     As shown in this last section, the fact (2) is guaranteed by the
     Feynman relation Eqn.\ref{fey} which relates the roton minimum
     to the peak of the structure factor.
      Across the phase boundary $ p=p_{c1} $ in Fig.2, the resulting
      solid could be an in-commensurate solid ( IC-NS )with vacancies or interstitials even at $ T=0
      $ whose condensation leads to  to the
      formation of the SS-v and SS-i respectively where $ \langle \psi \rangle \neq 0 $ \cite{and,ches,ander}.
      There is still some remanent of the SF in the IC-NS, the supersolid phase does exist as an equilibrium
      ground state.
      In this section, we assume $  | g | $  is sufficiently large in Fig.2 and study the SF
      to the SS transition across $ p=p_{c1} $.
      Because $ \psi $ is {\em also critical } through the transition, we can not simply
      integrate it out like in the last section.
      In fact, in the effective GL theory, we have to treat $ n $ and
      $ \psi $ on the same footing.
      From Eqn.\ref{sl} and  Eqn.\ref{sfnormal}, we can see
      that $ n $ and $ \psi $ have very similar propagators,
      so the lattice formation in $ n
      $ sector with  $ n(x)= n_{0}+ \sum^{\prime}_{\vec{G}} n_{\vec{G}} e^{i \vec{G} \cdot
       \vec{x} } $ where $ G=k_{n} $ and the superfluid density wave ( SFDW ) formation in $ \psi $ sector
      with  $ \langle \psi(\vec{x} ) \rangle  = \psi_{0}+ e^{i \theta_{2} } \sum^{P}_{m=1}
      \Delta_{m} e^{i  \vec{Q}_{m}  \cdot \vec{x} } $ where $ Q_{m}= k_{r}=k_{n} \sim 2\pi/a \sim 2\pi/3.17 \AA $
      have to happen simultaneously.
      From Hansen-Verlet criterion\cite{tom}, when $ S(k_{n} )/n_{0} $ is
      sufficiently large, solidification
      in the $ n $ sector occurs, so the roton minimum remains {\em finite} just before its condensation.
      The SFDW $ \rho=|\psi |^{2} $ is simply locked to ( or commensurate with ) the underlying normal
      solid ( $ n $ ) lattice. In fact, this locking is dictated by
      the $ \rho $ density- $ \delta n $ density couplings in Eqn.\ref{int}.
      If the coupling  $ g $ is attractive $ g_{v} < 0 $, the SFDW just
      coincides with the $ n $ lattice. If it is repulsive
      $ g_{i} > 0 $, then it simply
      shifts the SFDW by suitable constants along the three unit vectors in
      the direct lattice. These constants will be determined in the
      next section for different  $ n $ lattices.
      Namely, the supersolid states consist of two inter-penetrating lattices formed by
      the $ n $ lattice and the $\psi_{2} $ superfluid density wave.
      In fact, in a carefully prepared super-pressured
      sample, the roton minimum survives up to as high pressure as
      $ p_{r} \sim 120 \ bar $. This fact suggests the roton
      minimum in the meta-stable state in a super-pressured sample
      is replaced by a SFDW which is commensurate with the $ n $ lattice in the stable
      SS state. Strikingly, this pressure $ p_{r}\sim 120 \ bar $ is close to $
      p_{c2} \sim 170 \ bar $ in Fig.2 where the NCRI was extrapolated to vanish
      in the PSU's experiments \cite{chan}. This consistency also
      lead some support to the above picture. Obviously, when the
      pressure is so high  that $ p > p_{c2} $ in Fig.2, C-NS is the only possible
      ground state, any remanent of the SF completely disappears. So
      the vertical axis in the Fig.2 shows the SF-SS-NS series as the pressure  $ P $ increases at $ T=0 $
      Combining the roton condensation picture \cite{ring} in the last section with the
      results reviewed in the introduction,
      we can sketch the following phase diagram Fig.2 of the complete QGL Eqns. \ref{sl}, \ref{sfnormal} and
      \ref{int}.

\vspace{0.25cm}

\begin{figure}
\includegraphics[width=8cm]{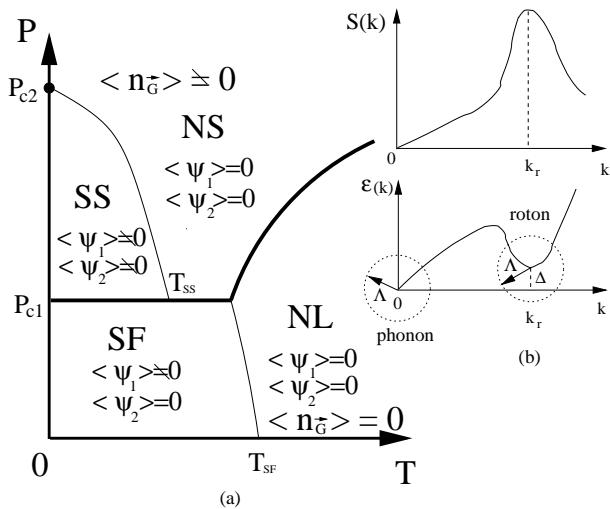}
\caption{ (a)  The theoretical phase
   diagram of GL model Eqns.\ref{sl},\ref{sfnormal} and \ref{int} in the pressure $ P $ versus temperature
   $ T $ plane.  This phase diagram only happens
   when $ |g| $ is sufficiently large ( Fig.3 ).  The SS is likely to be the vacancies induced supersolid ( SS-v ).
   $ T_{SS} $ is an effective measurement of the strength of the coupling constant $
   g \neq 0 $ in Eqn.\ref{int}.
   The $ \langle \psi_{1} \rangle $ and $\langle \psi_2  \rangle $ stand for the condensations in zero momentum and the first reciprocal lattice vectors
   respectively ( See Eqns.\ref{ssi} and \ref{ssv} ).
   Thick ( thin ) lines are 1st ( 2nd ) order transitions.
   The SF to the SS transition is a simultaneous combination of the SFDW transition in $ \psi $ sector
     driven by the roton condensation  at $ k_{0}=k_{r} $  and
    the NS transition in the $ n $ sector driven by the divergence of the structure function
    at $ k_{0}=k_{r}=k_{n} $.
   (b)  The structure factor and the separation  of low ( phonon ) and high ( roton ) momenta
   regime in the SF.}
\label{fig2}
\end{figure}

\vspace{0.25cm}

    As can be seen from the Fig.2, starting from the SF side, as the pressure is increased at a given
   temperature, there are two
   possible states (1) At  very low $
    T < T_{SS} $, the  Bose-Einstein condensation ( BEC)  of $ \psi_{2} $
    leads to the supersolid state where $ \langle \psi_{1} \rangle \neq 0, \langle \psi_{2} \rangle \neq 0, \langle n_{\vec{G}} \rangle
    \neq 0 $. The instability happens at a finite wavevector $ k=k_{r} $ instead of  at $ k=0 $.
    (2)  At higher temperature $ T_{SS} < T < T_{SF} $, there is a direct SF to NS transition.
    The SS state is certainly
    different from a conventional normal solid phase where
    $ \langle \psi_{1} \rangle =  \langle \psi_{2} \rangle = 0, \langle n_{\vec{G}} \rangle \neq 0 $.
    In addition to  the conventional translational
    and rotational orders characterized by  by $ n $,
    the SS also has the ODLRO  characterized by $ \psi_{1} $ and $ \psi_{2} $.
    When decreasing the temperature at a given pressure, if $ p < p_{c} $, the NL becomes SF at $ T=T_{SF} $,
    the instability happens at $ k=0 $.
    If $ p > p_{c} $, the NL becomes a NS first at $ T=T_{m} $, then there is a
    SFDW onset transition from the NS to a SS phase at $ T=T_{SS} $.

    Because the SF to SS transition driven by the roton condensation can be either weakly or strong first
    order, in principle, Eqn.\ref{sfdensity} works precisely only in the SF side, it is not easy
    to study the precise nature of the SS state from the SF side. It turns out that it is more convenient to
    study the properties of the SS state from the NS side in the next section.

\section{ The normal solid to the supersolid transition }
     Starting from the NS side, we assume that
     the local tunneling processes in the NS-PH only leads to local fluctuations of $ \psi $ with {\em a
     gap} $ \Delta (p) $ in Fig.3, so there is no long-range phase coherence and no supersolid in this case
     ( $ T_{SS} = 0 $ ).
     Taking the NS-PH state as the reference state, we will show that
     if the coupling $ g $ in Eqn.\ref{int} is sufficiently negative ( or  positive  ),
     the supersolids SS-v or SS-i can be realized by
     adding small number of vacancies  to NS-v or interstitials to the
     NS-i ( Fig.3). The attractive ( repulsive ) interaction $ g $ is shown to be crucial to
     raise the normal solid to the vacancy ( interstitials ) induced supersolid transition
     temperature $ T_{SS-v} $ ( $ T_{SS-i} $ )  above the zero
     temperature in the Fig.3.  In fact, the temperature $ T_{SS} $
     becomes an effective measure of the coupling strength $  g $.
     We find that SS-v is more likely than SS-i.
     However, in order to be complete, we study both cases on the same footing.
     It is also constructive to compare SS-v with SS-i even though
     the SS-i is unlikely to be relevant to the Helium 4 system.
     When the two kinds of SS show
     different properties, we treat them differently, when they share
     the same properties, we treat them just in the same notation SS.
     The SS phase naturally and consistently fits into the phase diagram.

   In the NS, $ \psi $ stands for the vacancies or interstitials coming from the large zero point quantum
   fluctuations \cite{and,ches,ander}. Inside the NS, the translational symmetry is already broken, we
   can simply set $ \delta  n( \vec{x} )=   n( \vec{x} )-n_{0} = \sum^{\prime}_{\vec{G}} n_{\vec{G}} e^{i \vec{G} \cdot
   \vec{x} } $ and put it into Eqn.\ref{int} to look at the effects
   of the coupling constants $ g $ and $ v $.
   Imagine that at a given pressure $ p > p_{c1} \sim 25 \ bar $, if tuning $ g \rightarrow 0
   $, but keeping $ v $ intact, then the normal solid becomes an asymptotically P-H symmetric
   normal solid ( NS-PH ) in Fig.3.
   Inside the NL, the mass  of $ \psi $ at $ \vec{k}=0 $ is $ t=T-T_{XY} $ which sets up the
   temperature scale of the problem. Inside the NS-PH, it is easy to see
   that the mass of $ \psi $ at $ \vec{k}=0 $ is $ t_{NS-PH} = t+ v \sum^{\prime}_{\vec{G}} | n( \vec{G}
   )|^{2} $. If we take the temperature scale $ t=T-T_{XY} $ as the reference
   scale, then $ t_{NS-PH} =T + \Delta (p)  $ where $  \Delta(p)= v  \sum^{\prime}_{\vec{G}} | n( \vec{G}  )|^{2}
   -T_{XY} >0 $ is the $ T=0 $ gap for the local superfluid mode $
   \psi $ at $ \vec{k}=0 $ in the NS-PH. Because as the pressure $ p $ increases,
   the repulsive interaction $ v $ also increases, so it is reasonable to assume that
   $ \Delta ( p ) $ is a  monotonic increasing function of
   $ p $. Because it is a first order transition across $ p_{c1} $, just like the roton gap $ \Delta_{r} > 0 $
   remains finite just before the first order transition, $ \Delta ( p )
   $ also remains finite just after the first order transition, namely, $ \Delta( p^{+}_{c1} ) > 0 $( Fig.3 ).

\vspace{0.25cm}

\begin{figure}
\includegraphics[width=8cm]{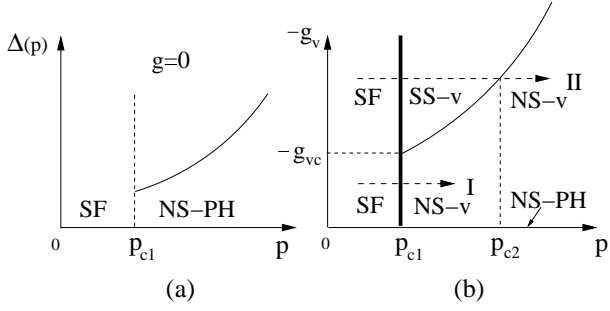}
\caption{ (a)  The gap $ \Delta(p) $  of the local fluctuating
superfluid mode $ \psi $ in the P-H symmetric commensurate normal
solid ( NS-PH ) which exists only at $ g=0 $. We assume it is a
monotonic increasing function of the pressure $ p $. (b) The zero
temperature phase diagram of $ g_{v} $ verse the pressure $ p $. The
NS-PH only exists at $ g_{v}=0 $.  Any $ g_{v} \neq 0 $ will
transfer the NS-PH into the NS-v. If the experimental is along the
path I, then it is a direct 1st order SF to NS-v transition in
Fig.1, if it is along the path II, it is a 1st order SF to SS-v,
then  a 2nd order SS-v to NS-v transition in Fig.2. } \label{fig3}
\end{figure}

\vspace{0.25cm}

   Taking this NS-PH as a reference state, we then gradually turn on
   $ g $ and see how the ground state evolves.
   In  the presence of the periodic potential of $ n(x) $ lattice,
   $ \psi $ will form a Bloch wave. In principle, a full energy
   band calculation incorporating the interaction $ u $  is
   necessary to get the energy bands of $ \psi $. Fortunately,
   qualitatively important physical picture of GL Eqns.\ref{sl}, \ref{sfnormal}, \ref{int}
   can be achieved without such a detailed energy band calculation.
   We expect that the self interaction $ u $ only leads to some
   renormalization of the parameters $ t $
   and $ K $.  Writing the $ g $ term in momentum space $ g \sum_{\vec{G}} n(
   \vec{G} ) \psi^{*}(\vec{G}) \psi(0) $ where $ \vec{G} \neq 0 $
   and integrating out $ \psi(\vec{G}) $, we get
   a perturbative expansion on the eigen-energy $  \epsilon_{\mu}(0) $
   at the origin $ \vec{K}=0 $, it is easy to see the $ n's $ term
   is of the form $ -(-g)^{n} $. Up to the third order $ g^{3}_{\mu} $, the series is:
\begin{eqnarray}
   \epsilon_{\mu} (0) & = & t_{NS-PH}
   - g^{2}_{\mu} P \frac{ n^{2}(\vec{G}) }{ K_{\mu } G^{2} }
   \nonumber  \\
   &  +  & g^{3}_{\mu}
   \sum^{\prime}_{\vec{G}_{1}}  \sum^{\prime}_{\vec{G}_{2}}
   \frac{  n(\vec{G}_{1}) n(\vec{G}_{2}) n(-\vec{G}_{1}-\vec{G}_{2} )}
   { K_{\mu } G^{2}_{1} K_{\mu } G^{2}_{2} }  +  \cdots
\end{eqnarray}
  where  the term linear in $ g $ vanishes, $ \mu =v,i $ stands for
  vacancies and interstitials respectively and the shortest
  reciprocal lattice vector is
    $ G= 2 \pi/a $ with  $ a \sim  3.17 \AA $ the lattice constant of the solid $ ^{4} He $,
    the $ n(\vec{K})= e^{-K^{2} a^{2}/4 \alpha } $  where $ K $ is any reciprocal lattice vector
    can be taken as a Gaussian where $ \alpha $ is the width of the Gaussian.

   For vacancies, $ g_{v} < 0 $, without writing out the coefficients explicitly, the expansion is
\begin{equation}
  \epsilon_{v}(0)= t_{NS-PH}-g^{2}_{v}-|g_{v}|^{3}-|g_{v}|^{4}- \cdots
\end{equation}
   so the coefficient has {\em the same } sign. Assuming the series
   converges, for  any $  g_{v}  < 0  $, we can write $  \epsilon_{v}
   (0)= t-f_{v} ( g_{v} ) $  where the $  f_{v} ( g_{v} ) \geq 0, f_{v} ( 0 )=0 $
   is a monotonic increasing function of  $ g_{v} $ and likely has
   no upper bound.

   For interstitials, $ g_{i} > 0 $, without writing out the coefficients
   explicitly, the expansion is
\begin{equation}
  \epsilon_{i} (0)= t_{NS-PH}-g^{2}_{i}+ g^{3}_{i}-g^{4}_{i}+ \cdots
\end{equation}
   so the coefficient has {\em oscillating } sign. Assuming the series
   converges, we can write $  \epsilon_{i}
   (0)= t-f_{i} ( g_{i} ) $ which holds for any $  g_{i}  $.  Because of the oscillating
   nature of the expansion coefficients, it is hard to judge the nature
   of the function of $ f_{i}( g_{i} ) $ except we know $ f_{i}( 0 ) = 0 $.
   The different expansion series of $ f_{v}( g_{v} ) $ and $ f_{i}( g_{i} ) $ indicate that quantum fluctuations
   may favor vacancies over interstitials.
   However, for simplicity, we assume $ f_{i}( g_{i} ) $ is also a monotonic increasing function of  $ g_{i}
   $, so we can discuss vacancies and interstitials induced supersolids on the same footing.
   In the following, we discuss SS-v and SS-i respectively.

\subsection{ Vacancies induced supersolid: SS-v }
    The mass of $ \psi_{v} $ was {\em decreased } to $ t_{\psi_{v}} = T +  \Delta( p ) - f_{v}( g_{v} ) =
   T-T_{SS-v} $ where  $ T_{SS-v} (p) =  f_{v}( g_{v} ) - \Delta(p) $ ( Fig.3).
   Because $ f_{v}( g_{v} ) $ is a monotonic increasing function of $ |g_{v} | $ and
   $ f_{v}( 0 ) = 0 $, defining a critical value $ f_{v}( g_{vc} )= \Delta ( p_{c1} ) $,
   then when  $ | g_{v} | < | g_{vc} | $, $ f_{v}( g_{v} ) <  \Delta ( p_{c1} ) $,
   the $ \psi_{v} $ mode remains massive, namely $ \langle \psi_{v} \rangle = 0 $.
   The C-NS remains to be the ground state even at $ T=0 $.
   It is important to stress that even the solid is a C-NS, it still does not have
   the P-H symmetry. For $ g_{v} < 0 $, the C-NS is a vacancy-like C-NS
   ( named as NS-v ) where the vacancy excitation energy
    $ \epsilon_{v} $ is lower than that of the interstitial $ \epsilon_{i} $ ( Fig.3b ).

   If $ | g_{v} | > | g_{vc} |  $, then
   $ T_{SS-v} ( p_{c1} )= f_{v}( g_{v} ) - \Delta( p_{c1} ) $ is raised above the zero
   temperature, the SS-v state exists in the Fig.3b. $ T_{SS-v} ( p_{c1}
   ) $ is also proportional to the superfluid density measured in
   the experiments.
   The resulting solid is an in-commensurate solid with vacancies
   even at $ T=0 $ whose condensation leads to $ \langle \psi_v \rangle \neq 0
   $. Of course, the number of vacancies $ n_{v} $ is quite small. The SS-v state
   has a lower energy than the NS-v state at sufficiently low temperature.
   As the pressure increases to $ p_{c2} $, $ T_{SS-v}( p_{c2} )= f_{v}( g_{v} ) - \Delta( p_{c2}
   ) =0 $ ( Fig.3 ). Then $  f_{v}( g_{v} ) = \Delta( p_{c2})  $, so
   $ T_{SS-v} (p) =  \Delta( p_{c2}) - \Delta(p) $ which
   becomes an effective experimental measure of the energy gap $ \Delta(p) $ in Fig.3a.
   Indeed, if reflecting the $ \Delta(p) $ in Fig.3a with respect to the
   horizontal axis and then shift by $ \Delta( p_{c2})=  f_{v}( g_{v}
   ) $, then one recovers the $ T_{SS-v} (p) $ in Fig.2.
   In the following, substituting the ansatz $ \langle \psi_{1}(\vec{x} )
   \rangle= a e^{i\theta_{1} } $ and $  \langle \psi_{2}(\vec{x} ) \rangle  = e^{i \theta_{2} } \sum^{P}_{m=1}
   \Delta_{m} e^{i  \vec{Q}_{m}  \cdot \vec{x} } $ where $ Q_{m}= Q $ into Eqn.\ref{int},
   we study the effects of  $ n $ lattice  on $ \psi= \psi_{1}+ \psi_{2} $.
   From  Eqn.\ref{int}, we can see $ n(x) $ acts as a periodic
   potential on $ \psi $.
   In order to get the lowest energy ground state, we must consider
   the following 4 conditions: (1)  because any complex $ \psi $ ( up to a global phase ) will lead to local
   supercurrents which is costy, so we can take $ \psi $ to be
   real, so $ \vec{Q}_{m} $ have to be paired as anti-nodal points. $ P $ has to be even (2)
   as shown from the Feynman relation Eqn.\ref{fey}, $ \vec{Q}_{m}, m=1,\cdots,P $ are
   simply $ P $ shortest reciprocal lattice vectors, then translational symmetry of the lattice dictates that
   $ \epsilon( \vec{K}=0 ) = \epsilon( \vec{K}=\vec{Q}_{m}) $,
   $ \psi_{1} $ and $ \psi_{2} $ have to condense at the same time (3) The point group symmetry of the lattice
   dictates $ \Delta_{m} =\Delta $ and is real
   (4) for the vacancies case, the interaction is attractive $ g_{v} < 0
   $, from Eqn.\ref{int}, the SFDW-v simply sits on top of the $ n $ lattice,
   so the Superfluid Density wave $ \rho=|\psi|^{2} $ simple sits on top of the $ n $ lattice as much as
   possible. This is reasonable, because vacancies are hopping on
   near the lattice sites. From Eqn.\ref{int}, the attractive interaction also favors
   $ \psi( x = \vec{R}/2 ) \sim 0 $ where $  \vec{R}/2 $ stands for
   any interstitial sites which are in the middle of lattice points.
   It turns out that the the 4 conditions can fix the relative phase and magnitude of $ \psi_{1} $ and
   $\psi_{2} $ to be $ \theta_{2}=\theta_{1}, \Delta= a/P $, namely:
\begin{equation}
    \psi_{ss-v} = \psi_{0} ( 1 + \frac{2}{P} \sum^{P/2}_{m=1} \cos \vec{Q}_{m} \cdot \vec{x} )
\label{ssv}
\end{equation}
   where $ \psi_{0} = a e^{i \theta} $ depends on the temperature and pressure.
   Note that in contrast to a uniform superfluid, the magnitude of $
   \psi $ is changing in space.
   This field satisfies the Bloch theorem with the crystal momentum
   $ \vec{k}=0 $ and  the Fourier components are $ \psi( \vec{K}=0)=
   a, \psi( \vec{K}= \vec{Q}_{m} )= a/P $. They have the same sign and decay in magnitude. In principle, higher
   Fourier components may also exist, but they decay very rapidly, so
   can be neglected without affecting the physics qualitatively.

\begin{figure}
\includegraphics[width=6cm]{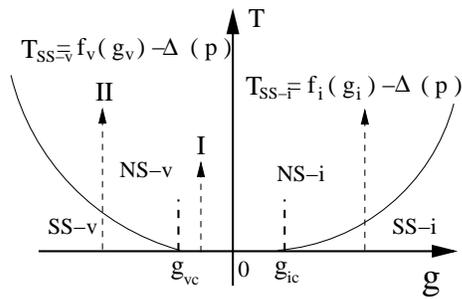}
\caption{ The phase diagram of $ T $ versus $ g $ at a given
 pressure $ p_{c1} < p < p_{c2} $. The finite temperature
 transitions denoted by the dashed line II at a given $  | g_{v} | > | g_{vc} | $
  ( or $ g_{i} > g_{ic}  $  ) is in the 3D $ XY $ universality class described by Eqn.\ref{3dxy}.
   However, if $ | g_{v} | < | g_{vc} | $, then the ground state at $ T=0 $ is a
   vacancy-like C-NS( named as NS-v ).
   Similar thing can be defined for an interstitial-like C-NS ( named as NS-i ) when $ g_{i} < g_{ic} $. }
\label{fig4}
\end{figure}

\subsection{ Interstitial induced supersolid: SS-i }
   The discussion is quite similar to the vacancy case except that
   (1) if $ g_{i}< g_{ic} $, the C-NS is an interstitial-like C-NS ( named as NS-i ) where the interstitial excitation
   energy $ \epsilon_{i} $ is lower than that of the vacancy $ \epsilon_{v}
   $ (2) if $ g_{i} > g_{ic} $, the resulting solid is an in-commensurate solid with
   interstitials even at $ T=0 $ whose condensation leads to $ \langle \psi_i \rangle \neq 0 $,
   the SS-i state exists in the Fig.2.
   The arguments to determine the lattice structure of the SS-i goes the same as
   those in the vacancies case except the condition (4), for the
   SS-v, the interaction is attractive $ g_{v} < 0
   $, the SFDW-v simply sits on top of the $ n $ lattice. However, for the
   interstitials case, the interaction is repulsive $ g_{i} > 0
   $ favors $ \psi(x=0) \sim 0 $,
   so the Superfluid Density wave $ \rho=|\psi|^{2} $ can avoid the $ n $ lattice as much as
   possible. This is reasonable, due to the competing of the the two orders, the superfluid component
   emerges from the places where the  normal solid  component is suppressed .
   It turns out that the the 4 conditions can fix the relative phase and magnitude of $ \psi_{1} $ and
   $\psi_{2} $ to be $ \theta_{2}=\theta_{1}+ \pi, \Delta= a/P $,
   namely:
\begin{equation}
    \psi_{ss-i}= \psi_{0} ( 1- \frac{2}{P} \sum^{P/2}_{m=1} \cos \vec{Q}_{m} \cdot \vec{x} )
\label{ssi}
\end{equation}
   where $ \psi_{0} = a e^{i \theta} $ depends on the temperature and pressure.
   Note that the crucial sign difference from the vacancies case
   which make an important difference from the X-ray scattering from SS-v and
   SS-i to be discussed in the next section.
   Again in contrast to a uniform superfluid, the magnitude of $
   \psi $ is changing in space.
   This field satisfies the Bloch theorem with the crystal momentum
   $ \vec{k}=0 $ and  the Fourier components are $ \psi( \vec{K}=0)=
   a, \psi( \vec{K}= \vec{Q}_{m} )= - a/P $. They oscillate in sign and decay in magnitude. In principle, higher
   Fourier components may also exist, but they decay very rapidly, so
   can be neglected without affecting the physics qualitatively.

   In the following, we will discuss different lattice structures of
   $ n $ when $ P= 2,4,6,8,12 $ respectively. Obviously, the Superfluid Density wave $ \rho=|\psi|^{2} $
     has the same Bravais lattice structure as the the $ n $
     lattice. However, as shown in the following, even $ n $ is a
     Bravais lattice, the SFDW may be the same Bravais lattice plus a
     few basis.

   (a) P=2: $ \vec{Q}_{1} = -\vec{Q}_{2}= \vec{Q} $ are a pair of anti-nodal points.
   They are the two shortest reciprocal lattice vectors generating a 1
   dimensional lattice embedded in a 3 dimensional system. The field
   is $ \psi ( \vec{x} ) =a(1- \cos \vec{Q} \cdot \vec{x} ) $.
   There is a superfluid density wave formation transition inside the normal
   solid which is a 2nd order transition in the universality class of 3D XY model( Fig.2).
   The local SFDW operator $ \rho^{l}_{s} = |  \psi(\vec{x} ) |^{2}=
   a^{2} ( 1- \cos \vec{Q} \cdot \vec{x})^{2} $. It breaks translational invariance only along
   1-dimension which is similar to Smectic-A or Smectic-C phase in the liquid crystal \cite{tom}.
   The maxima of the SFDW  $ \psi_{max}=2 a $   appear exactly in the middle of lattice
   points  at $ \vec{a}= \frac{1}{2} \vec{a}_{1} $. So the SFDW forms the dual lattice of the 1d lattice which is
   also a 1d lattice.

   (b) P=4: $ \vec{Q}_{3} = - \vec{Q}_{1}, \vec{Q}_{4}= - \vec{Q}_{2}, \vec{Q}_{1} \cdot \vec{Q}_{2}=0 $,
   $ \vec{Q}_{i}, i=1,2,3,4 $ form the 4 corners of a square. They are the four shortest
   reciprocal lattice vectors generating a 2 dimensional square lattice embedded in a 3 dimensional system.
   The field is $ \psi( \vec{x})=a[1- \frac{1}{2}( \cos \vec{Q}_{1} \cdot \vec{x} + \cos \vec{Q}_{2} \cdot \vec{x}  )] $,
   The local SFDW operator $ \rho^{l}_{s} = |  \psi(\vec{x} )
   |^{2} $. The maxima of the SFDW  $ \psi_{max}=2 a $ appear exactly in the middle of lattice
   points at $ \vec{a}= \frac{1}{2} ( \vec{a}_{1} + \vec{a}_{2}) $.
   So the SFDW forms the dual lattice of the square lattice which is
   also a square lattice.

    (c) P=6: there are two cases need to be discussed separately.
    (c1) $ \vec{Q}_{i}, i=1,2,3,4,5,6 $ form the 6 corners of a
    hexagon. They consist of the 6 shortest
    reciprocal lattice vectors generating a 2 dimensional triangular lattice embedded in a 3 dimensional system.
    The field is $ \psi ( \vec{x})=a[1- \frac{1}{3}( \cos \vec{Q}_{1} \cdot \vec{x} + \cos \vec{Q}_{2} \cdot \vec{x}
     +  \cos \vec{Q}_{3} \cdot \vec{x}  )] $.
    The maxima of the SFDW  $ \psi_{max}= 3/2 a $ appear in the middle of lattice
   points at $ \vec{a}= \pm \frac{1}{3} ( \vec{a}_{1} + \vec{a}_{2}
   ) $. They form the {\em dual} lattice of the triangular lattice which is
   a honeycomb lattice. The honeycomb lattice is not a Bravais
   lattice which has two triangular sublattices $ A $ and $ B $,
   it can be considered as one triangular lattice $ A $ plus a
   basis. This can be understood intuitively: there are two
   equivalent ways to shift the $ n $ lattice, one way to get the
   sublattice $ A $, the other to get the sublattice $ B $. Putting
   $ A $ and $ B $ together forms the  SFDW which takes the honeycomb lattice
    (c2) $ \vec{Q}_{i}, i=1,2,3,4,5,6 $ are the 6 shortest
   reciprocal lattice vectors generating a cubic lattice. The maxima of the SFDW $ \psi_{max}=2 a $ appear exactly in the middle of lattice
   points at the 8 points $ \vec{a}=  \frac{1}{2} ( \pm \vec{a}_{1} \pm
   \vec{a}_{2} \pm \vec{a}_{3} ) $. So the SFDW forms the dual lattice of the cubic lattice which is
   also a cubic lattice.

   (d)  $ P=8 $:  $ \vec{Q}_{i}, i=1,\cdots,8 $ form  the 8 shortest
   reciprocal lattice vectors generating a $ bcc $ reciprocal
   lattice which corresponds to a $ fcc $ direct lattice. The field is
   $ \psi ( \vec{x})=a[1- \frac{1}{4}( \cos \vec{Q}_{1} \cdot \vec{x} + \cos \vec{Q}_{2} \cdot \vec{x}
   +  \cos \vec{Q}_{3} \cdot \vec{x} +  \cos \vec{Q}_{4} \cdot \vec{x}  )]
   $. The maxima of the SFDW $ \psi_{max}=2 a $ appear in all the edge centers such as $ (1/2,0,0) $ etc. and
   the centers of any cube such as $ (1/2,1/2,1/2) $ etc.  It is easy
   to see these points can be achieved by simply shifting the $ n $
   lattice by $ (1/2,1/2,1/2) $, so the SFDW also forms a $ fcc $ direct lattice.

 (e)  $ P=12 $:  $ \vec{Q}_{i}, i=1,\cdots,12 $ form  the 12 shortest
   reciprocal lattice vectors generating a $ fcc $ reciprocal
   lattice which corresponds to a $ bcc $ direct lattice. The field is
    $ \psi( \vec{x}) = a[1- \frac{1}{6}( \cos \vec{Q}_{1} \cdot \vec{x} + \cos \vec{Q}_{2} \cdot \vec{x}
     +  \cos \vec{Q}_{3} \cdot \vec{x} +  \cos \vec{Q}_{4} \cdot \vec{x}+
      \cos \vec{Q}_{5} \cdot \vec{x} +  \cos \vec{Q}_{6} \cdot \vec{x}   )]
      $.
   The maxima of the SFDW  $ \psi_{max}= 4/3 a $ appear along
   any square surrounding the center of the cube such as $ (1/2,\beta,0)
   $ or $ (1/2,0,\gamma) $ etc. In fact, one can achieve the SFDW lattice by
   shifting the the center of $ bcc $ to the 3 face centers, so all
   these points are on the edge centers and face centers which are only discrete points on
   the square surrounding the center of the cube. We expect
   the continuous whole square is due to the artifact of the
   approximation $ \psi( x=0 ) \sim 0 $ imposed.
   So the SFDW
   $ ^{4}He $ in Vycor glass takes a $ bcc $ lattice.

   Unfortunately, a spherical
   $ k = Q $ surface can not lead to lattices with different lengths
   of primitive reciprocal lattice vectors such as a $ hcp $ lattice. This is
   similar to the classical liquid-solid transition described by Eqn.\ref{sl} where a single
   maximum peak in the static structure factor can not lead to a $ hcp $ lattice \cite{tom}.
   Another difficulty with the $ hcp $ lattice is that the $ hcp $
   lattice is not a Bravais lattice, it consist of two inter-penetrating simple
   hexagonal lattices shifted by $ \vec{a}=  \frac{1}{3} \vec{a}_{1}
   + \frac{1}{3} \vec{a}_{2} + \frac{1}{2} \vec{a}_{3}  $.
   Here we can simply take the experimental fact that $ n $ forms a $ hcp
   $ lattice without knowing how to produce such a lattice from a GL
   theory Eqn.\ref{sl}. Despite the technical
   difficulty, because for an idea $ hcp $ lattice $ c/a =\sqrt{8/3}
   $, an $ hcp $ lattice has 12 nearest neighbors, so its local
   environment may resemble that of an $ fcc $ lattice.
   We expect the physics ( except the anisotropy of NCRI in the $ hcp $ lattice to be discussed
   in the next two sections ) is qualitatively the same as that in $ fcc
   $ direct lattice.  From the insights achieved from the other
   lattices, one can achieve the SFDW lattice by shifting the $ hcp $
   lattice of $ n $ by  $ \vec{a}=  \frac{2}{3}  ( \vec{a}_{1}
   +  \vec{a}_{2} ) + \frac{1}{4} \vec{a}_{3}  $.

\subsection{ Discussions on both SS-v and SS-i}
   In fact,  for both SS-v and SS-i, we can write $ n $ and $ \psi $ sector in a more
   symmetric way: $ n( \vec{x}
     )=  n_{0}  + \sum^{\prime}_{\vec{G}} n_{\vec{G}} e^{i \vec{G} \cdot
     \vec{x} },  \psi ( \vec{x}
     )=  \psi_{0}  + \sum^{\prime}_{\vec{G}} \psi_{\vec{G}} e^{i \vec{G} \cdot
     \vec{x} } $. It is easy to see that in the SS, the
     translational symmetry  $ \vec{x} \rightarrow \vec{x} + \vec{a},
     n_{\vec{G}} \rightarrow n_{\vec{G}} e^{i \vec{G} \cdot
     \vec{a} },  \psi_{\vec{G}} \rightarrow \psi_{\vec{G}} e^{i \vec{G} \cdot
     \vec{a} }  $ is broken down to $ \vec{a} = \vec{R},
     n_{\vec{G}} \rightarrow n_{\vec{G}} e^{i \vec{G} \cdot
     \vec{R} } = n_{\vec{G}}, \psi_{\vec{G}} \rightarrow \psi_{\vec{G}} e^{i \vec{G} \cdot
     \vec{R} }=  \psi_{\vec{G}} $.
   In the Fig.2, in the NL side, as the temperature is lowered, the
   symmetry breaking happens in $ \psi_{1} $ at $ k=0 $, the NL gets into the SF. However, as shown in this section,
   in the NS side, the symmetry breaking happens in both the $ \psi_{1} $ at $ k=0 $ and  $ \psi_{2} $ sector at $ P $
   discrete points  simultaneously,
   the NS gets to the SS state at a much lower critical temperature $ T_{SS} $. The results achieved  from the NS
   side in this section indeed confirm Fig.2 achieved from the roton condensation
   picture in the SF phase in the last section.

\section{  The zero temperature transition from NS to SS driven by the pressure  }

    The analysis so far focused on finite temperature and
   mean field level.
   Some interesting physics near the {\em finite temperature} NS to SS transition in Fig.1
   was explored in \cite{dor} by considering the coupling of elastic degree of freedoms to the SF mode.
   For example, the sound velocity will acquire a dip similar to the
   specific heat cusp in the $ \lambda $ transition in superfludi
   Helium. In this and following sections,
   I will push the Ginsburg-Landau (GL) theory in the previous sections to zero
   temperature and to include all the possible low energy
   fluctuations above the mean filed solutions achieved in
   the previous sections. Particularly, I will work out the new
   elementary low energy excitations including vortex loop excitations in a SS and study how they defer
   from the low energy excitations in solids and superfluids. In
   principle, if these elementary low energy excitations can be detected by X-ray scattering, neutron scattering,
   acoustic wave attenuation and heat capacity experiments in solid Helium 4 can prove or disprove the existence
   of the supersolid in Helium 4. In practice, the detection may still be
   complicated by sample quality. No matter if a supersolid indeed
   exists in Helium 4, these results should be interesting in its
   own and may have application in other systems.

      So far, we only look at the mean field solutions corresponding to vacancies and interstitials. Here we
      discuss excitations above the mean field solutions. It turns
      out that the excitations in both cases are the same, so we
      discuss both cases at same time.
      In the SS-v and SS-i, the wavefunctions can be written as
\begin{equation}
       \psi_{ss}= \psi_{0} ( 1 \pm \frac{2}{P} \sum^{P/2}_{m=1} \cos\vec{Q}_{m} \cdot \vec{x}
       ),~~~~~ \psi_{0} = | \psi_{0}| e^{i \theta}
\label{ssvi}
\end{equation}
      where $ \pm $ sign corresponds to SS-v and SS-i respectively.
      Obviously, there are also
      topological defects in the phase winding  of $ \theta
      $ which are  vortices.  At $ T \ll T_{SS} $, the vortices  can only appear in tightly bound
      pairs. However, as $ T \rightarrow T^{-}_{SS} $, the vortices start to
      become liberated, this process renders the total NCRI to vanish above $ T > T_{SS}
      $. In addition to the superfluid  $ \theta $ mode in SS states, there
   are also lattice phonon modes $ \vec{u} $ in both $ n $ sector and $ \psi_{2} $ sector.
   However, it is easy to see that
   the coupling Eqn.\ref{int} is invariant under $ \vec{x}
   \rightarrow \vec{x} + \vec{u}, n( \vec{G} ) \rightarrow n(
   \vec{G} ) e^{ i \vec{G} \cdot \vec{u} },  \psi(
   \vec{G} ) \rightarrow  \psi( \vec{G} ) e^{ i \vec{G} \cdot \vec{u}
   } $, so the lattice
   phonon modes  in $ \psi $  are locked to those in the conventional $ n $ lattice.
   This is expected because there is only one kind of translational
   symmetry breaking, therefore only one kind of lattice phonons.
  Inside the NS side, the translational symmetry is already broken,
  so we can parameterize the density deviation order parameter
  $ \delta n (\vec{x}, \tau ) = n( \vec{x}, \tau ) - n_{0}  $ and the SF complex
  order parameter $  \psi ( \vec{x}, \tau ) $  as:
\begin{eqnarray}
 \delta n( \vec{x}, \tau ) & = &  \sum^{\prime}_{\vec{G}} n_{\vec{G}} e^{i \vec{G} \cdot
     ( \vec{x} + \vec{u}( \vec{x}, \tau ) ) }   \nonumber  \\
 \psi ( \vec{x}, \tau ) & = &  \psi_{0} ( \vec{x}, \tau ) [  1  \pm \frac{1}{P} \sum^{\prime}_{\vec{G}}
      e^{i \vec{G} \cdot
     ( \vec{x} + \vec{u}( \vec{x}, \tau ) )} ]
\label{orderhe}
\end{eqnarray}
     where the $ \psi_{0}( \vec{x}, \tau )= | \psi_{0}( \vec{x}, \tau ) | e^{ i \theta(  \vec{x}, \tau ) }  $
     is the SF order parameter, $ \vec{u}( \vec{x}, \tau )  $ are the 3 lattice phonon modes,
     the $ \pm $ means vacancy or interstitials induced supersolids respectively,
     $ n^{*}_{\vec{G}}= n_{-\vec{G}} $ the " $\prime $ " means the sum over the shortest non-zero
     reciprocal lattice vector  $ \vec{G} $ and $ P $ is the number of them.
     From Eqn.\ref{orderhe}, we can identify the SS density order parameter $ \rho_{\vec{G}}(\vec{x}, \tau
     )=  e^{i \vec{G} \cdot \vec{u}( \vec{x}, \tau ) } $.
     The effective action to describe the NS to SF transition at $
     T=0 $ consistent with all the lattice symmetries  and the global
     $ U(1) $ symmetry is:
\begin{eqnarray}
   {\cal L}  & = & \psi^{\dagger}_{0} \partial_{\tau} \psi_{0}  +
    c_{\alpha \beta} \partial_{\alpha} \psi^{\dagger}_{0} \partial_{\beta} \psi_{0} + r | \psi_{0} |^{2} +
    g  |\psi_{0} |^{4}          \nonumber  \\
    & + &  \frac{1}{2} \rho_{n}  ( \partial_{\tau} u_{\alpha} )^{2} +
    \frac{1}{2} \lambda_{\alpha \beta \gamma \delta} u_{\alpha \beta
    } u_{\gamma \delta }     \nonumber  \\
    & + &  a^{0}_{\alpha \beta }  u_{\alpha \beta } \psi^{\dagger}_{0} \partial_{\tau} \psi_{0}
     + a^{1}_{\alpha \beta } u_{\alpha \beta } | \psi_{0} |^{2} + \cdots
\label{quan}
\end{eqnarray}
    where  $ r=p-p_{c2} $ with $ p_{c2} \sim 170 \ bar $ ( Fig.5),
    $ \rho_{n} $ is the normal density, $ u_{\alpha \beta }= \frac{1}{2}( \partial_{\alpha}
    u_{\beta} + \partial_{\beta} u_{\alpha} ) $ is the strain
    tensor, $ \lambda_{\alpha \beta \gamma \delta} $ are the bare
    stress tensor dictated by the symmetry of the
    lattice, it has 5 (2) independent elastic constants for a hcp  ( isotropic )
    lattice. For a uniaxial lattice such as $ hcp $ lattice, all the coefficients $ c_{\alpha
    \beta}, a^{0}_{\alpha \beta }, a^{1}_{\alpha \beta } $  take
    the same form $ c_{\alpha
    \beta}= c_{z} n_{\alpha} n_{\beta} + c_{\perp} (\delta_{\alpha
    \beta}- n_{\alpha} n_{\beta} ) $  where $ \vec{n} $ is a unit vector along the uni-axis \cite{elas,dor}.
    In the NS state $ r >0, \langle \psi_{0}( \vec{x}, \tau
     ) \rangle =0 $, the 3 lattice phonon modes $ \vec{u}( \vec{x}, \tau )  $ become the 3 ordinary
     ones. While inside the SS state $ r < 0,  \langle \psi_{0}( \vec{x}, \tau ) \rangle \neq 0 $.
     From the  parameterizations of  $ \psi ( \vec{x}, \tau ) $,
     we can see if the prefactor $ \langle \psi_{0}( \vec{x}, \tau ) \rangle \neq 0
     $, then $ \psi ( \vec{x}, \tau ) $ condenses at both $
     \vec{G}=0 $ and any other non-zero reciprocal lattice vectors $ \vec{G}
     $ to form the superfluid density wave ( SFDW ) $ \rho^{s}_{l}= |\psi( \vec{x}, \tau )|^{2} $
     inside the SS.
    The  $ a^{0}_{\alpha \beta } $ and $ a^{1}_{\alpha \beta } $
    couplings come from the original couplings $ \delta n ( \vec{x}, \tau )
     \psi^{\dagger} \partial_{\tau} \psi $ and $ \delta n ( \vec{x}, \tau ) |
     \psi |^{2}  $  respectively in the GL in Sect.I.
    If setting all the couplings between $ \psi_{0} $ and $ u_{\alpha} $
    vanish, the $ \psi_{0} $ sector describes the SF to Mott insulator transition
    in a {\em rigid } underlying lattice \cite{boson}.  So we will study how the NS to SS transition  at $ T=0, p= p_{c2}
    $ in a rigid lattice is affected by its coupling to a quantum fluctuating
    lattice.  Under the Renormalization  group ( RG ) transformation, $
    \tau^{\prime} = \tau/b^{z},  x^{\prime} = x /b $ and $ \psi^{\prime}
    = \psi/Z $. If we choose $ z=2, Z= b^{-d/2} $, the $ g^{\prime}=
    g b^{2-d} $. It is well known that the SF to Mott insulator transition
    in a {\em rigid } underlying lattice has the mean field
    exponents with $ z=2, \nu=1/2, \eta=0 $ at $ d \geq 2$\cite{boson}.
    We also choose  $ u^{\prime}_{\alpha}
    = u_{\alpha}/Z $, then $ \rho^{\prime}_{n}= b^{-2} \rho_{n} $,
    so the lattice phonon kinetic energy term is irrelevant near the QCP.
    It is easy to see $ a^{\prime}_{0} = b^{-d/2-1} a_{0} $, so $
    a_{0} $ is always irrelevant.  $ a^{\prime}_{1} = b^{1-d/2} a_{1}
    $, so both $ g $ and $ a_{1} $'s upper critical dimension is $
    d_{u}=2 $, so, in principle, a $ \epsilon=2-d $ expansion is
    possible for both $ g $ and $ a_{1} $. However, both are irrelevant at $ d=3 $. We conclude that NS
    to SS transition at $ T=0 $ remains the same as that in a rigid
    lattice described by
\begin{equation}
   {\cal L}_{T=0}= \psi^{\dagger}_{0} \partial_{\tau} \psi_{0}  +
    c_{\alpha \beta} \partial_{\alpha} \psi^{\dagger}_{0} \partial_{\beta} \psi_{0} + r | \psi_{0} |^{2} +
    g  |\psi_{0} |^{4} + \cdots
\label{z2}
\end{equation}
     Namely, it is a transition with mean field exponents $ z=2, \nu=1/2,
     \eta=0 $.

\begin{figure}
\includegraphics[width=6cm]{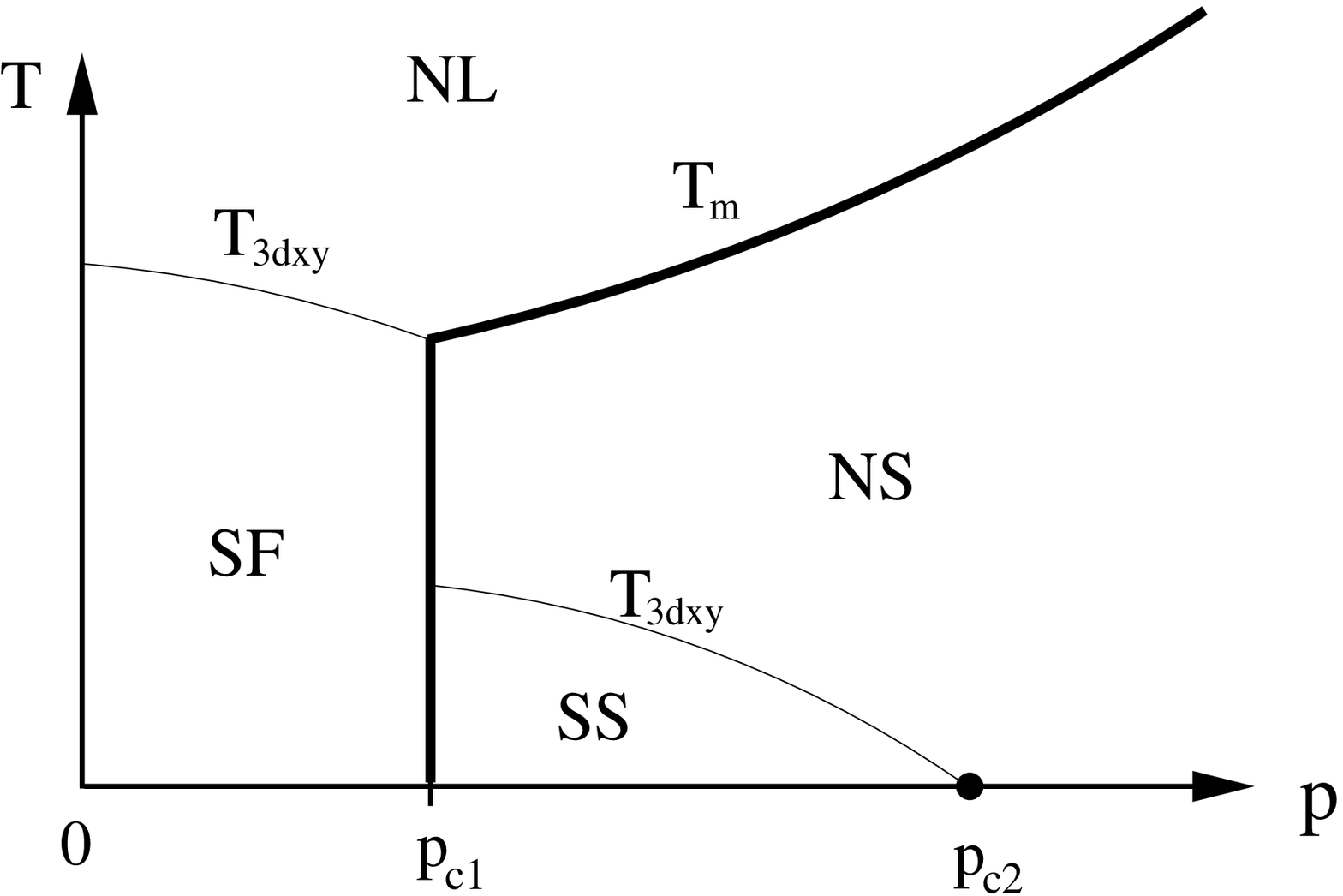}
\caption{ Possible temperature $ T $ versus pressure $ p $ phase
diagram of Helium 4. SF is the superfluid phase, SS is the
supersolid phase, NS is the normal solid phase, NL is the normal
liquid phase. $ T_{3dxy} $ is the 3d XY transition. $ T_{m} $ is the
1st order melting transition. The dot is the zero temperature
transition from the NS to the SS which is a transition with mean
field exponents $ z=2, \nu=1/2, \eta=0 $. Thick (thin) line is 1st
(2nd) order transition. This figure can be obtained from Fig.2 by
rotating $ 90^{\circ} $. }
 \label{fig5}
\end{figure}

    If neglecting the $ \tau $ dependence by setting
    $ u_{\alpha}( \vec{x}, \tau )= u_{\alpha}( \vec{x} ), \psi_{0}( \vec{x}, \tau ) =   \psi_{0}( \vec{x} ) $,
    then Eqn.\ref{quan} reduces to the classical action studied in \cite{elas,dor}.
    For the classical case, $ x^{\prime} = x /b, \psi^{\prime}
    = \psi/Z, u^{\prime}_{\alpha} = u_{\alpha}/Z  $, if we choose $
    Z=b^{(2-d)/2} $, then $  g^{\prime}=
    g b^{4-d}, a^{\prime}_{1} = b^{2-d/2} a_{1} $, so both $ g $ and $ a_{1} $'s upper critical dimension is $
    d_{u}=4 $. So in principle, a $ \epsilon=4-d $ expansion is
    possible for both $ g $ and $ a_{1} $, the putting $ \epsilon=1
    $ for $ d=3 $. In \cite{elas}, it was shown that due to the specific heat exponent
    of the 3d XY  model $ \alpha=-0.012< 0 $,  the $ a^{1}_{\alpha \beta } $
    coupling is irrelevant, so the NS to SS transition remains to be
    a classical 3d XY transition at finite temperature given by:
\begin{equation}
  {\cal L}_{T}  =  K_{NS} | \nabla \psi_{0} |^{2} + t_{NS} | \psi_{0} |^{2} +  u_{NS}  |\psi_{0} |^{4}  + \cdots
\label{3dxy}
\end{equation}
      where $ t_{NS}= T-T_{SS} $ and  $ T_{SS-v} (p) =  f_{v}( g_{v} ) -
      \Delta(p) = \Delta( p_{c2} )- \Delta(p) $
      and $ T_{SS-i} (p) =  f_{i}( g_{i} ) - \Delta(p)=  \Delta ( p_{c2} )- \Delta(p)  $.
      Obviously, it is still a 3d XY transition. It was shown in
      \cite{dor} that the coupling to the phonon mode $ \vec{u} $
      will not change the universality class at finite temperature.
      As the pressure increased to $ p_{c2} $, $ T_{SS-v} $ or $
      T_{SS-i} $ are suppressed to zero, the system becomes a C-NS
      where $ \langle \psi \rangle =0 $. It is important to stress that even
      at $ p > p_{c2} $, the solid is a C-NS, it still does not have
      the P-H symmetry because $ g \neq 0 $. For $ g_{v} < 0 $, the
      C-NS is vacancy like ( NS-v ) where the vacancy excitation energy is lower than that of the interstitial.
      For $ g_{i} > 0 $, the
      C-NS is interstitial like ( NS-i ) where the interstitial excitation energy is lower than that of the vacancy.

\section{ The NCRI of SF and SS states at $ T=0 $ and $ T > 0 $.}


   In order to calculate the superfluid density $ \rho_{s} $ explicitly, we need to look at
   how the system's free energy responses to a fictitious gauge potential $
   \vec{A} $. We find that when $ \Delta \gg \Delta_{c} $, $ \rho_{s}( T=0 ) = \rho = \int d^{d} x |
   \psi( \vec{x}, \tau ) |^{2} = \int d^{d} x  ( | \psi_{1}( \vec{x}, \tau )
   |^{2} + | \psi_{2}( \vec{x}, \tau )|^{2} ) $ where the crossing terms between $ \psi_{1} $ and $ \psi_{2} $
   drop out due to  the momenta conservation.

   In the SF state, at low $ T $, the quantum fluctuations induced by the pressure are
   important. Let's first look at the quantum phase fluctuations.
   The phase fluctuation action is given by $ {\cal L}_{p} = \frac{1}{2 g } \frac{1}{\beta} \sum_{i \omega_{n} }
   \int \frac{d^{d} k}{ (2 \pi)^{d} } ( \omega^{2}_{n} + k^{2} ) |
   \theta( \vec{k}, \omega_{n} ) |^{2} $ where $ g=\frac{1}{
   \rho_{s} } $ controls the strength of quantum phase fluctuations and
   the superfluid phonon velocity has been set equal to 1 for
   simplicity. It is easy to see that at $ T=0 $, $ \langle \theta^{2}( \vec{x}, \tau) \rangle_{T=0} $ is infra-red ( IR )
   finite, so the quantum phase fluctuations alone will not lead to any instability.
   However, it will lead to superfluid density depletion even
   before reaching the phase boundary of SF to SS transition in Fig.1 and Fig.2,
   although the depletion may be quite small. This fact explains why $ T_{SF}(p) $ bends to the left
   slightly as the pressure $ p $ increases. At finite $ T $, the thermal fluctuations $ \langle \theta^{2}( \vec{x}, \tau)
   \rangle_{T}- \langle \theta^{2}( \vec{x}, \tau) \rangle_{T=0} \sim T^{d-1} $  lead to $ \rho_{s}( T )=  \rho_{s}( T=0 ) -
    c T^{2} $ at $ d=3 $. It is well known the superfluid density $
    \rho(T) \sim \rho(T=0)- a T^{4} $, while the Bose condensation density $ n_{b}(T)
    \sim n_{b}(T=0)-b T^{2} $. So strictly speaking, $ \psi $ sector can
    only describe the Bose condensation. This is expected, because
    the $ n $ sector in the SF phase also contributes to the
    superfluid density, but not to the BEC.

   Then let's look at the roton fluctuations.
   Setting the roton gap $ \Delta_{2} = \Delta^{2} $, at $ T=0 $, the quantum roton
   fluctuations $ \langle \phi^{2}_{2} ( \vec{x}, \tau) \rangle_{T=0} \sim \log \frac{\Lambda}{
   \Delta } $ where $ \Lambda $ is the upper cutoff is  IR
   logarithmic divergent as $ \Delta \rightarrow 0 $ which signifies the instability to the lattice formation.
   Due to this IR divergence, the 1st order SF to SS transition may happen well before $ \Delta $ becomes zero, namely, at $ \Delta= \Delta_{c} > 0
   $. This is consistent with the picture described in Sec.III. At finite $ T $, the thermal roton fluctuations
   $ \langle \phi^{2}_{2}( \vec{x}, \tau) \rangle_{T}- \langle \phi^{2}_{2}( \vec{x}, \tau) \rangle_{T=0} \sim  ( \log \frac{\Lambda }{ \Delta } ) e^{-\frac{\Delta}{T} } $ when $ T \ll \Delta \ll \Lambda $.

   In the SS-v and SS-i states, the $ n(x) $ forms a lattice, at the same time, the unstable roton part is replaced by a
   stable SFDW formation commensurate with the underlying $ n $ lattice.
   Obviously, the $ n(x) $ normal lattice takes away the vast majority of density from the superfluid density
   even at $ T=0 $. It can be shown that superfluid density
   from the $ \psi_{1} $ sector is  isotropic $ \rho_{1} \sim K a^{2}
   $, while the  superfluid density from $ \psi_{2} $ sector turns out to
   anisotropic in $ hcp $ lattice
   $ \rho_{2,ij}  \sim  \sum^{P}_{m=1} |\Delta_{m} |^{2} Q_{mi} Q_{mj}/Q^{2}
   $ where $ \Delta_{m}= \pm a/P $ for SS-v or SS-i. Therefore, the total superfluid density in SS phase
   is:
\begin{equation}
    \rho^{SS}_{ij}  \sim  \rho_{1} \delta_{ij} + \rho_{2ij} \sim  a^{2} ( K \delta_{ij}
   +\frac{1}{P^{2}} \sum^{P}_{m=1} Q_{mi} Q_{mj}/Q^{2} )
\end{equation}

    Taking $ P=6 $, the anisotropy is quite small. Solid $^{4}He$ in a bulk takes
    a $ hcp $ lattice with $ c/a \sim 1.63 $ which is quite close to the idea value $ c/a=\sqrt{8/3} $.
    The three primitive reciprocal lattice vectors are $
    G_{1}= G_{2}= \frac{ 4 \pi }{ \sqrt{3} a }, G_{3}= \frac{2 \pi}{c} $.
    We can estimate the anisotropy of the NCRI in the $ hcp $
    lattice. If the rotation axis is along the $ c $ axis, the NCRI
    is $  \rho_{11} \sim K a^{2} + v_{r}  \sum^{6}_{m=1} |\Delta_{m} |^{2} Q_{m1}
    Q_{m1}/Q^{2} $
    If the rotation axis is along the $ a $ ( or $ b $ )  axis, the NCRI
    is $ \rho_{33} \sim K a^{2} + v_{r}
    \sum^{6}_{m=1} |\Delta_{m} |^{2} Q_{m3} Q_{m3}/Q^{2} $.
    The anisotropy mainly comes from the $ \psi_{2} $ sector.
    Setting $ s=G_{1}/G_{3} $, for the idea value $ s= \frac{4
    \sqrt{2}}{3} > 1 $, so $ \rho_{11} > \rho_{33} $. Namely, the NCRI response is
    larger when one  is rotating the sample around the $ c $ axis than that when
    one is rotating the sample around the $ a $ or $ b $ axis.
    However, as the pressure is increased, $ s $ decreases, the
    anisotropy of the NCRI also decreases. In the PSU experiments,
    the samples are poly-crystal, the relative orientation of the
    rotation axis to the $ c $ axis is not known, so it's hard to
    test this prediction with poly-crystals.

\section{ Non-topological elementary excitations and spectral weights in the SS }

   Classical non-equilibrium hydrodynamics in SS was investigated for a long
   time \cite{and,hydro}. These hydrodynamics will break down at very low temperature where quantum
   fluctuations dominate. However, the quantum nature of the excitations in
   the SS has not been studied yet. Here,
   we will study the quantum characteristics of low energy excitations in the SS.
   For example, how the phonon spectra in the SS differ from that
   in a NS and how the SF mode in the SS differs from that in
   a SF. Inside the SS,  $  \langle \psi_{0}( \vec{x}, \tau ) \rangle = a $, we can
    write $  \psi_{0}( \vec{x}, \tau )= \sqrt{a+ \delta \rho} e^{ i
    \theta (\vec{x}, \tau ) } $ and plug it into the Eqn.\ref{quan}:
\begin{eqnarray}
   {\cal L}  &  = &  i \delta \rho \partial_{\tau}
   \theta + \rho^{s}_{\alpha \beta} \partial_{\alpha} \theta \partial_{\beta} \theta
   + \frac{1}{2} \delta \rho   S^{-1}_{0}    \delta \rho
      \nonumber \\
   & + &  \frac{1}{2}[ \rho_{n}  ( \partial_{\tau} u_{\alpha} )^{2} +
     \lambda_{\alpha \beta \gamma \delta} u_{\alpha \beta
    } u_{\gamma \delta } ]   \nonumber  \\
    &  + &  a^{0}_{\alpha \beta }  u_{\alpha \beta } i  \partial_{\tau} \theta
     + a^{1}_{\alpha \beta } u_{\alpha \beta } \delta \rho  +  \cdots
\label{ss1}
\end{eqnarray}
     where we already dropped  $  i \partial_{\tau} \theta $ term which is
     irrelevant inside the SS phase ( although it is very important
     in describing the $ T=0 $  NS to SS transition discussed in the last section
     ), $ S_{0} $ is the bare density-density correlation function.
      Integrating out the massive magnitude $ \delta \rho $
     fluctuations, we get:
\begin{eqnarray}
   {\cal L}  & = & -\frac{ S_{0} }{2} ( i  \partial_{\tau}
   \theta + a^{1}_{\alpha \beta } u_{\alpha \beta } )^{2} +
    \rho^{s}_{\alpha \beta} \partial_{\alpha} \theta \partial_{\beta}
    \theta      \nonumber  \\
    & +  &  \frac{1}{2}[  \rho_{n}  ( \partial_{\tau} u_{\alpha} )^{2} +
    \lambda_{\alpha \beta \gamma \delta} u_{\alpha \beta
    } u_{\gamma \delta }]    \nonumber    \\
    &  +  & a^{0}_{\alpha \beta }  u_{\alpha \beta } i  a \partial_{\tau} \theta + \cdots
\label{ss2}
\end{eqnarray}
     Expanding the square, we get the
     effective action describing the low energy
     modes inside the SS phase:
\begin{eqnarray}
   {\cal L}  & = &   \frac{1}{2} [ \kappa ( \partial_{\tau} \theta )^{2} +
    \rho^{s}_{\alpha \beta} \partial_{\alpha} \theta \partial_{\beta}
    \theta]            \nonumber \\
    &  + &  \frac{1}{2}[ \rho_{n}  ( \partial_{\tau} u_{\alpha} )^{2} +
    \lambda_{\alpha \beta \gamma \delta} u_{\alpha \beta
    } u_{\gamma \delta } ] + a_{\alpha \beta }  u_{\alpha \beta } i  \partial_{\tau} \theta
\label{ss3}
\end{eqnarray}
    where  $ \kappa $ is the SF compressibility defined in Eqn.\ref{xy}  and $ \rho^{s}_{\alpha \beta} $
    is the SF stiffness which has the same symmetry as $  a^{0}_{\alpha \beta }  $,
    $  a_{\alpha \beta }= a^{0}_{\alpha \beta } + S_{0}
    a^{1}_{\alpha \beta } $ where
    $ S_{0}(\vec{k},\omega ) $ is the bare {\em SF density} correlation function.
    Again, the $ \rho_n $ is the normal density, the $ \lambda_{\alpha \beta \gamma \delta} $ is the stress tensor.
    Obviously, the last term is the crucial {\sl Berry phase }
    coupling term which couples the lattice phonon modes to the
    SF mode. The factor of $ i $ is important in this coupling.
    By integration by parts, this term can also be written as $ a_{\alpha
    \beta} ( \partial_{\tau} u_{\beta} \partial_{\alpha} \theta
    + \partial_{\tau} u_{\alpha} \partial_{\beta} \theta ) $ which
    has the clear physical meaning of the coupling between the SF
    velocity $ \partial_{\alpha} \theta $ and the velocity of
    the lattice vibration $ \partial_{\tau} u_{\beta} $.
    It is this term which makes
    the low energy modes in
    the SS to have its own characteristics which could be detected by
    experiments. The invariance under the Galilean transformation \cite{son,yoo} dictates that $ a_{\alpha,\beta}=\rho_n \delta_{\alpha,\beta}- \rho^{s}_{\alpha,\beta} $.

    In this section, we neglect the topological
    vortex loop excitations in Eqn.\ref{ss1}. In the next section, we will
    discuss these vortex loop excitations in detail. In the
    following, we discuss two extreme cases: isotropic solid and $ hcp $ lattice
    separately. Usual samples are between the two extremes.

\subsection{ Isotropic solid }

    A truly isotropic solid can only be realized in a highly poly-crystalline
    sample. Usual samples are not completely isotropic. However, we
    expect the simple physics brought about in an isotropic solid
    may also apply qualitatively to other samples which is very
    poly-crystalline.

    For an isotropic solid, $ \lambda_{\alpha \beta \gamma \delta}=
    \lambda \delta_{\alpha \beta} \delta_{\gamma \delta}
    + \mu ( \delta_{\alpha \gamma } \delta_{\beta \delta}+ \delta_{\alpha \delta} \delta_{\beta \gamma
    } ) $ where $ \lambda $ and $ \mu $ are Lame coefficients, $
    \rho^{s}_{\alpha,\beta}= \rho^{s} \delta_{\alpha,\beta}, a_{\alpha,\beta}= a
    \delta_{\alpha,\beta} $ where  $ a= \rho_n- \rho_s $. In $ ( \vec{q}, \omega_{n} ) $ space, Eqn.\ref{ss1} becomes:
\begin{eqnarray}
    {\cal L}_{is} & = &  \frac{1}{2}[ \rho_{n} \omega^{2}_{n} + ( \lambda+2 \mu
    ) q^{2} ] |u_{l}( \vec{q},\omega_{n} ) |^{2}         \nonumber  \\
     & + & \frac{1}{2} [ \kappa \omega^{2}_{n} + \rho_{s} q^{2} ] |\theta ( \vec{q},\omega_{n} ) |^{2}
                    \nonumber  \\
     & + &  a q \omega_{n} u_{l}( -\vec{q}, - \omega_{n} ) \theta ( \vec{q},\omega_{n} )
                    \nonumber  \\
    & + &  \frac{1}{2}[ \rho_{n} \omega^{2}_{n} + \mu q^{2} ] |u_{t}( \vec{q},\omega_{n} ) |^{2}
\label{is}
\end{eqnarray}
     where $ u_{l}( \vec{q},\omega_{n} )= i q_{i} u_{i}(
     \vec{q},\omega_{n}/q $ is the longitudinal component,
     $ u_{t}( \vec{q},\omega_{n} )= i \epsilon_{ij} q_{i} u_{j}( \vec{q},\omega_{n}
     )/q $ are transverse components of the
     displacement field. Note that Eqn.\ref{is}
     shows that only longitudinal component couples to the
     superfluid  $ \theta $ mode, while the two transverse components
     are unaffected by the superfluid mode. This is expected, because
     the superfluid mode is a longitudinal density mode itself which
     does  not couple to the transverse modes.

     From Eqn.\ref{is}, we can identify the
     longitudinal-longitudinal phonon correlation function:
\begin{equation}
  \langle u_{l} u_{l} \rangle= \frac{ \kappa \omega^{2}_{n} + \rho_{s} q^{2} }{
  ( \kappa \omega^{2}_{n} + \rho_{s} q^{2} ) (  \rho_{n} \omega^{2}_{n} + ( \lambda+2 \mu
    ) q^{2} ) + a^{2} q^{2} \omega^{2}_{n} }
\end{equation}
     The $ \langle \theta \theta \rangle $ and $ \langle u_{l} \theta \rangle $ correlation
     functions can be similarly written down. By doing the
     analytical continuation $ i \omega_{n} \rightarrow \omega + i
     \delta $, we can identify the two poles of  all the correlation
     functions at $ \omega^{2}_{\pm}= v^{2}_{\pm} q^{2} $ where the
     two velocities $ v_{\pm} $ is given by $ v^{2}_{\pm}  = [ \kappa( \lambda+2 \mu ) + \rho_{s} \rho_{n} +
    a^{2} \pm     \sqrt{ ( \kappa( \lambda+2 \mu ) + \rho_{s} \rho_{n} +   a^{2} )^{2}- 4 \kappa \rho_{s} \rho_{n} ( \lambda+2 \mu ) } ]/ 2 \kappa \rho_{n} $.
    If setting $ a =0 $, then $ v^{2}_{\pm} $ reduces to the longitudinal phonon
    velocity $ v^{2}_{lp}=  ( \lambda + 2 \mu )/ \rho_{n} $ and
    the superfluid velocity $ v^{2}_{s} = \rho_{s}/\kappa $ respectively.
    Of course, the transverse phonon velocity $ v^{2}_{tp}= \mu /
    \rho_{n} $ is untouched. For notation simplicity, in the
    following, we just use $ v_{p} $ for $ v_{lp} $.
    Inside the SS, due to the very small superfluid density $
    \rho_{s} $, it is expected that $ v_{p} > v_{s} $.
    In fact, in isotropic solid $ He^{4} $, it was measured that $ v_{lp} \sim 450-500
    m/s, v_{t} \sim 230 \sim 320 m/s $ and $ v_{s} \sim 366 m/s $ near the melting curve \cite{melt}.
    It is easy to show that $ v_{+} > v_{p}> v_{s} > v_{-} $ and
    $ v^{2}_{+} + v^{2}_{-}  >  v^{2}_{p} + v^{2}_{s} $, but  $ v_{+} v_{-} = v_{p} v_{s} $,
    so $ v_{+} + v_{-} > v_{p} + v_{s} $ ( see
    Fig.1 ). Note that because the Galilean invariance dictates $ a= \rho_n- \rho_s $,
    for $ \rho_s \ll \rho_n $, one can see $ \rho_{s} \rho_{n} +   a^{2} \gg \rho_{s} \rho_{n} $, so
    $ v_{+} $ ( $ v_{-} $ ) are considerably above ( below ) $ v_{p} $ ( $ v_{s} $ ), so the two supersolidons,
    especially the softening of the lower branch, may be
    easily distinguished by possible neutron scattering experiments.

\begin{figure}
\includegraphics[width=4cm]{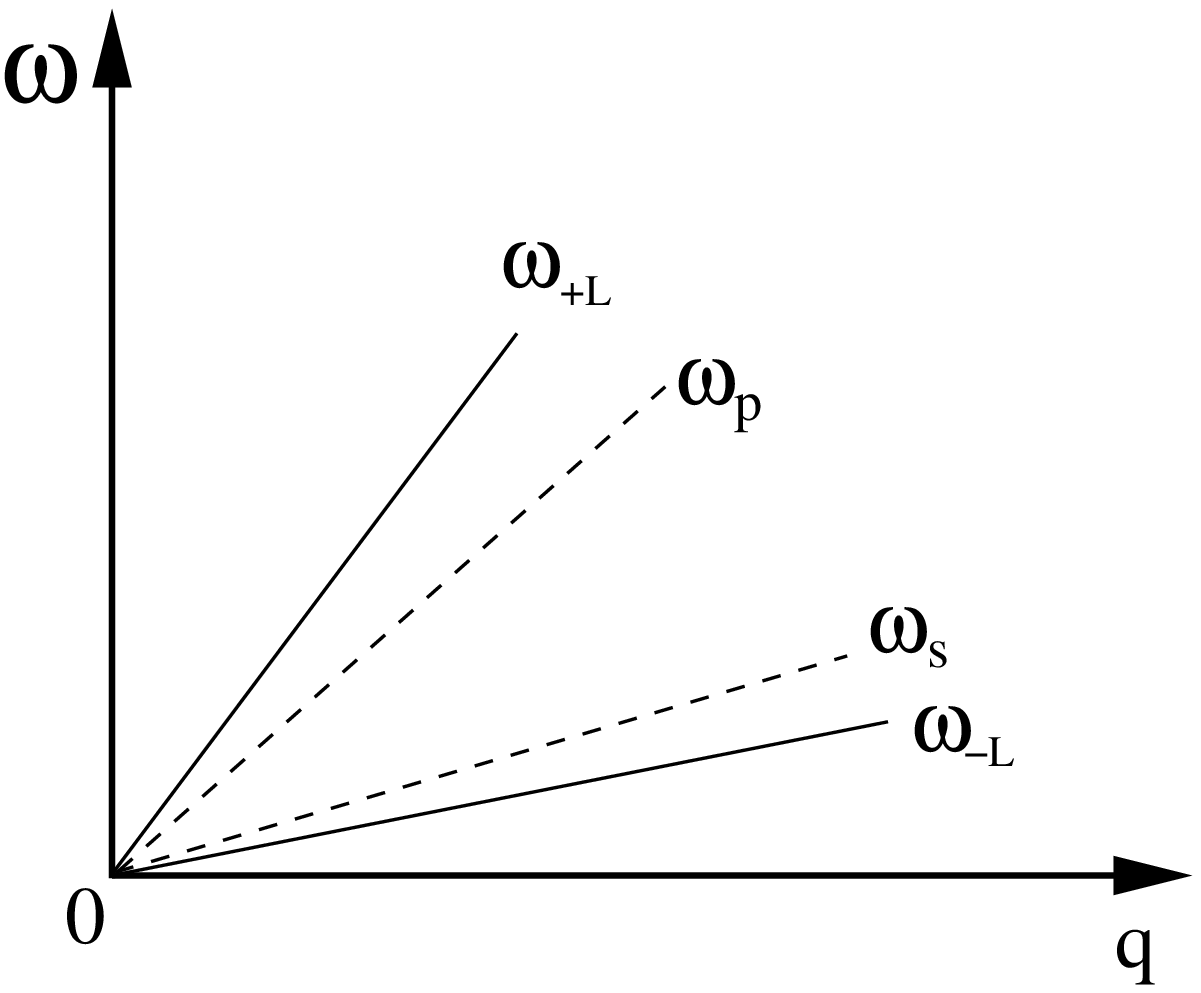}
\hspace{0.5cm}
\includegraphics[width=3cm]{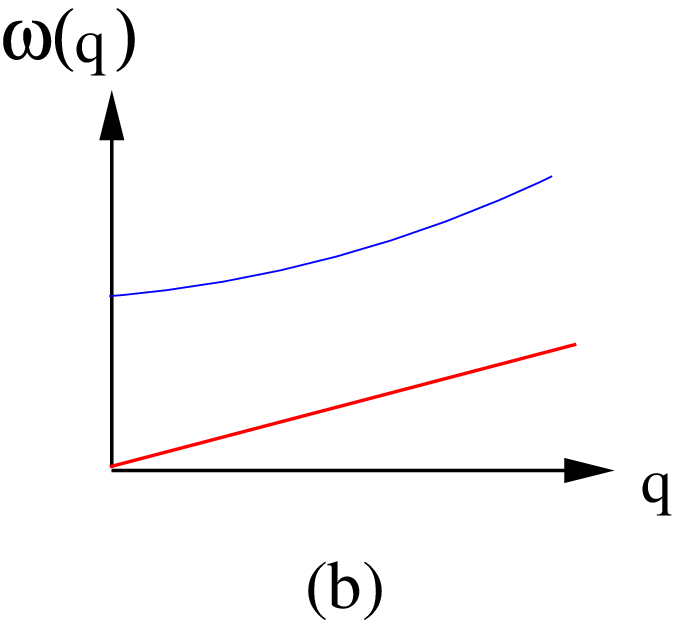}
\caption{ (a) The elementary low energy excitations inside a supersolid.
The coupling between the phonon mode $ \omega_{p} = v_{p}q $ ( the
upper dashed line ) and the superfluid mode $ \omega_{s}=v_{s}q $ (
the lower dashed line ) leads to the two new "supersolidon " modes $
\omega_{\pm}=v_{\pm} q $ ( solid lines ) in the SS. Their corresponding spectral weights are  listed in Eqn.\ref{uu0} and \ref{tt}.
These two new supersolid modes and the spectral weights should be detected by in-elastic neutron
scattertings. (b) The excitations inside a lattice CDW supersolid discussed in the appendix B.
It has one gapped mode with a CDW gap and a gapless superfluid mode.
The corresponding spectral weights are worked out in \cite{bragg}. } \label{fig6}
\end{figure}

    By doing the  analytical continuation $ i \omega_{n} \rightarrow \omega + i
     \delta $, we can take the imaginary part and find:
\begin{eqnarray}
   &  & Im  \langle u_{l} u_{l} \rangle_{i \omega_{n} \rightarrow \omega + i
     \delta}     \nonumber   \\
     & = & \frac{ v^{2}_{s}- v^{2}_{+} }{  v^{2}_{+}- v^{2}_{-} } \frac{ \pi}{ 2 \rho_{n} v_{+} } \frac{1}{q}
     [ \delta( \omega- v_{+}q ) -  \delta( \omega+ v_{+}q ) ]   \nonumber  \\
     & - & \frac{ v^{2}_{s}- v^{2}_{-} }{  v^{2}_{+}- v^{2}_{-} } \frac{ \pi}{ 2 \rho_{n} v_{-} } \frac{1}{q}
     [ \delta( \omega- v_{-}q ) -  \delta( \omega+ v_{-}q ) ]
\label{uu0}
\end{eqnarray}
    It is easy to see that the second term can be achieved from the first term just by $ v_{+} \leftrightarrow v_{-} $.
    Setting $ a=0 $, then $ v_{+} = v_{p} $, $ v_{-}=v_{s} $, the second term just vanishes, the first term recovers
    the excitation spectrum of the lattice phonons. When $ a \neq 0 $, then Eqn.\ref{uu0} becomes a mixing of the lattice phonons and superfluid
    phonons, the first and second term give the excitation energies and the two corresponding spectral weights.

    Very similarly, we can find
\begin{eqnarray}
   &  & Im  \langle \theta \theta \rangle_{i \omega_{n} \rightarrow \omega + i
     \delta}     \nonumber   \\
     & = & \frac{ v^{2}_{p}- v^{2}_{+} }{  v^{2}_{+}- v^{2}_{-} } \frac{ \pi}{ 2 \kappa v_{+} } \frac{1}{q}
     [ \delta( \omega- v_{+}q ) -  \delta( \omega+ v_{+}q ) ]   \nonumber  \\
     & - & \frac{ v^{2}_{p}- v^{2}_{-} }{  v^{2}_{+}- v^{2}_{-} } \frac{ \pi}{ 2 \kappa v_{-} } \frac{1}{q}
     [ \delta( \omega- v_{-}q ) -  \delta( \omega+ v_{-}q ) ]
\label{tt}
\end{eqnarray}
    It is easy to see that the second term can be achieved from the first term just by $ v_{+} \leftrightarrow v_{-} $.
    Setting $ a=0 $, then $ v_{+} = v_{p} $, $ v_{-}=v_{s} $, the first term just vanishes, the second term recovers
    the excitation spectrum of the superfluid phonons. When $ a \neq 0 $, then Eqn.\ref{tt} become a mixing of the lattice phonons and superfluid
    phonons, the first and second term give give the excitation energies and the two corresponding spectral weights.

    The two longitudinal modes in the SS can be understood from an
    intuitive picture: inside the NS, it was argued in \cite{van} that
    there must be a diffusion mode of vacancies in the NS. Inside
    the SS, the vacancies condense and lead to the extra superfluid mode.
    So the diffusion mode in the NS is replaced by the SF mode in
    the SS. The analysis in this subsection is similar to the atom-photon polaritons of atoms inside a cavity \cite{qedbec,phasediffusion}
    or exciton-photon polaritons of  excitons in quasi-two dimensional quantum wells inside a planar cavity \cite{expo}.

\subsection{ $ hcp $ crystal }

  Usual single $ hcp $ crystal samples may also contain dislocations,
  grain boundaries. Here we ignore these line and plane defects
  and assume that there are only vacancies whose condensation leads
  to the superfluid density wave inside the supersolid discussed in Sect.IV.

  For a uni-axial crystal such as an $ hcp $ lattice, the action is:
\begin{eqnarray}
   {\cal L}_{hcp}  & = &  \frac{1}{2} [ \rho_{n}  ( \partial_{\tau} u_{\alpha} )^{2} +
   K_{11} ( u^{2}_{xx}+u^{2}_{yy} ) + 2 K_{12} u_{xx} u_{yy}
   \nonumber  \\
    &  + &  K_{33} u^{2}_{zz} + 2 K_{13} ( u_{xx}+u_{yy} ) u_{zz}
    \nonumber  \\
    &  + &  2 ( K_{11}-K_{12} ) u^{2}_{xy} + K_{44} ( u^{2}_{yz} +
      u^{2}_{xz} ) ]   \nonumber  \\
   & +  &  \frac{1}{2}[ \kappa ( \partial_{\tau} \theta )^{2} +
    \rho^{s}_{z} ( \partial_{z} \theta )^{2}+  \rho^{s}_{\perp}
    ( ( \partial_{x} \theta )^{2}+ ( \partial_{y} \theta )^{2})   ]
     \nonumber  \\
   & + & [ a_{z} \partial_{z} u_{z} + a_{\perp}
   ( \partial_{x} u_{x} + \partial_{y} u_{y})  ] i  \partial_{\tau} \theta
\label{hcp}
\end{eqnarray}

    If $ \vec{q} $ is along $ \hat{z} $ direction, namely $ q_{z}
    \neq 0, q_{x}=q_{y}=0 $, then Eqn.\ref{hcp}  simplifies to:
\begin{eqnarray}
    {\cal L}^{z}_{hcp} & = &  \frac{1}{2}[ \rho_{n} \omega^{2}_{n} +
    K_{33} q^{2}_{z} ] |u_{z}( q_{z}, \omega_{n} ) |^{2}         \nonumber  \\
     & + & \frac{1}{2} [ \kappa \omega^{2}_{n} + \rho^{s}_{z} q^{2}_{z} ] |\theta ( q_{z},\omega_{n} ) |^{2}
                    \nonumber  \\
     & - & i a_{z} q_{z} \omega_{n} u_{z}( -q_{z}, - \omega_{n} ) \theta ( q_{z},\omega_{n} )
                    \nonumber  \\
    & + &  \frac{1}{2}[ \rho_{n} \omega^{2}_{n} + K_{44} q^{2}_{z}/4 ]|u_{t}( q_{z}, \omega_{n} ) |^{2}
\label{hcpz}
\end{eqnarray}
     where $ |u_{t}( q_{z}, \omega_{n} ) |^{2}= |u_{x}( q_{z},\omega_{n} ) |^{2} +  |u_{y}( q_{z},\omega_{n} ) |^{2} $
     stand for the two transverse modes with the velocity  $ v^{2}_{t}= K_{44}/4 \rho_{n} $. The superfluid mode only couples to the longitudinal $ u_{z} $
     mode, while the two transverse modes $ u_{x}, u_{y} $ are decoupled.
     Eqn.\ref{hcpz} is identical to Eqn.\ref{is} after the replacement $ u_{z}
     \rightarrow u_{l}, K_{33} \rightarrow \lambda + 2 \mu, a_{z}
     \rightarrow a $. It was found that $ v_{lp} \sim 540 m/s, v_{t}
     \sim 250 m/s $ when $ \vec{q} $ is along the $ \hat{z} $ direction \cite{sound}.
     Fig.2 follows after these replacements.

    Similarly, we can work out the action in the $ xy $ plane where
    $ q_{z}=0, q_{x} \neq 0, q_{y} \neq 0 $.
    Then $ u_{z} $ mode is decoupled, only $ u_{x}, u_{y} $ modes
    are coupled to the superfluid mode:
\begin{eqnarray}
    {\cal L}^{xy}_{hcp} & = &  \frac{1}{2}[
    \rho_{n}  ( \partial_{\tau} u_{\alpha} )^{2}
   + K_{11} ( u^{2}_{xx}+u^{2}_{yy} )
       \nonumber \\
   & + & 2 K_{12} u_{xx} u_{yy} + 2 ( K_{11}-K_{12} ) u^{2}_{xy} ]
     \nonumber  \\
   & +  &  \frac{1}{2}[ \kappa ( \partial_{\tau} \theta )^{2}
    + \rho^{s}_{\perp} ( \partial_{\alpha} \theta )^{2} ]
     \nonumber  \\
   & + &  a_{\perp} \partial_{\alpha} u_{\alpha} i  \partial_{\tau} \theta
      \nonumber  \\
   & + &  \frac{1}{2}[ \rho_{n}  ( \partial_{\tau} u_{z} )^{2}
      + K_{44}/4 ( \partial_{\alpha} u_{z} )^{2}]
\label{hcpxy}
\end{eqnarray}
     where $ \alpha, \beta =x,y $. By comparing Eqn.\ref{hcpxy} with
     Eqn.\ref{is}, we can see that $ K_{11} \rightarrow \lambda + 2 \mu,
     K_{12} \rightarrow \lambda $, so all the discussions in the isotropic case
     can be used here after the replacements.
     Fig.2 follows after these replacements.
     Namely, only the longitudinal component  in
     the $ xy $ plane is coupled to the $ \theta $ mode, while the
     transverse mode in the $ xy $ plane with velocity
     $ v^{2}_{txy}= ( K_{11}-K_{12})/2 \rho_{n} $ is decoupled. Obviously the transverse mode along
     $ \hat{z} $ direction $ u_{z} $ mode with the velocity $ v^{2}_{tz}= K_{44}/4 \rho_{n} $ is also decoupled.
     Note that the two transverse modes have different velocities.
     It was found that $ v_{lp} \sim 455 m/s, v_{tz}
     \sim 255 m/s, v_{txy} \sim 225 m/s $ when $ \vec{q} $ is along the $ xy $ plane \cite{sound}.

     Along any general direction $ \vec{q} $, strictly speaking, one can not even
     define longitudinal and transverse modes, so the general action
     Eqn.\ref{hcp} should be used \cite{tonyp}.
     Despite the much involved $ 4 \times 4 $ matrix
     diagonization in $ u_{x},u_{y},u_{z}, \theta $, we expect the qualitative physics is still
     described by Fig.2.

    In principle, inelastic neutron scattering experiments \cite{neutron}  or acoustic attenuation experiments \cite{acou}
    can be used to detect the predicted the low energy excitation spectra in the SS shown in Fig.2.

\section{ Topological elementary excitations in SS:  Vortex loops
and vortex  }

   In the last section, we studied the low energy excitations shown in the
   Fig.1 by neglecting the  topological vortex loop. Here, we will
   study how the vortex loop interaction in SS differ from that in
   the SF.   For simplicity,
   in the following, we only focus on the isotropic case. The
   formulations can be generalized to the $ hcp $ case
   straightforwardly. We can perform a duality transformation on
   Eqn.\ref{ss1} to the vortex loop representation:
\begin{equation}
   {\cal L}_{v} = \frac{1}{2 K_{\mu}} ( \epsilon_{\mu \nu \lambda \sigma
   } \partial_{\nu} a_{\lambda \sigma} - a \partial_{\alpha} u_{\alpha} \delta_{\mu \tau}  )^{2}
    + i 2 \pi a_{\mu \nu} j^{v}_{\mu \nu}
\label{dual}
\end{equation}
    where $ \mu, \nu, \lambda, \sigma  $ stand for space and time,
    but $ \alpha, \beta $ stand for the space components only, $ K_{0}= \kappa, K_{\alpha}= \rho_{s} $ and
    $  a_{\mu \nu}=- a_{\nu \mu} $ is an anti-symmetric tensor
    gauge field and $ j^{v}_{\mu \nu}= \frac{1}{2 \pi}  \epsilon_{\mu \nu \lambda \sigma
   } \partial_{\lambda } \partial_{\sigma } \theta $ is the anti-symmetric tensor vortex
   loop current due to the topological phase winding in $ \theta $.

    Eqn.\ref{dual} has the gauge invariance $a_{\mu \nu} \rightarrow
    a_{\mu \nu} + \partial_{\mu} \chi_{\nu}- \partial_{\nu}
    \chi_{\mu} $ where $ \chi_{\mu} $ is any 4-component field \cite{string}.
    It is the most convenient to choose the Coulomb gauge $ \partial_{\alpha} a_{\alpha
    \beta}=0 $ to get rid of the longitudinal component, then
    the transverse component is $ a_{t}= i \epsilon_{\alpha \beta
    \gamma} q_{\alpha} a_{\beta \gamma} /q $. It can be shown that $
    |a_{t}|^{2}=2 | a_{\alpha \beta} |^{2} $.
    Then Eqn.\ref{dual} is simplified to:
\begin{eqnarray}
 {\cal L}_{v} & = &  \frac{1}{2}[ \rho_{n} \omega^{2}_{n} + ( \lambda+2 \mu
     + a^{2}/\kappa ) q^{2} ] |u_{l}( \vec{q},\omega_{n} ) |^{2}        \nonumber  \\
     & + &  \frac{1}{2} ( q^{2}/\kappa + \omega^{2}_{n}/\rho_{s} ) |a_{t}|^{2}
     + \frac{2}{ \rho_{s} } q^{2} | a_{0 \alpha} |^{2}
                    \nonumber  \\
     & - &  a q^{2}/\kappa  u_{l}( -\vec{q}, - \omega_{n} ) a_{t}( \vec{q},\omega_{n} )
                    \nonumber  \\
    & + & i 2 \pi j^{v}_{ 0 \alpha } a_{0 \alpha} + i 2 \pi
    j^{v}_{\alpha \beta} a_{\alpha \beta }
\label{prop}
\end{eqnarray}
    where the transverse phonon mode $ u_{t} $ was omitted, because
    it stays the same as in the NS as shown in Eqn.\ref{is}.

   It is easy to see that only $ a_{t} $ has the dynamics, while $ a_{0
   \alpha} $ is static. This is expected, because although $ a_{\mu
   \nu} $ has 6 non-vanishing components, only the transverse component $ a_{t}
   $ has the dynamics which leads to the original gapless superfluid
   mode $ \omega^{2}= v^{2}_{s} q^{2} $. Eqn.\ref{prop} shows that
   the coupling is between the longitudinal phonon mode $ u_{l} $ and the
   transverse gauge mode $ a_{t} $. The vortex loop density is
   $ j^{v}_{ 0 \alpha }= \frac{ 1}{2 \pi} \epsilon_{\alpha \beta
    \gamma} \partial_{\beta}   \partial_{\gamma} \theta $ and the
    vortex current is $ j^{v}_{ \alpha  \beta }= \frac{ 1}{2 \pi} \epsilon_{\alpha \beta
    \gamma} [ \partial_{0},  \partial_{\gamma}] \theta $. Integrating out the $ a_{0 \alpha} $, we
    get the vortex loop density-density interaction:
\begin{equation}
    \pi \rho_{s} \int^{\beta}_{0} d \tau \int dx dy j^{v}_{ 0 \alpha
    }( \vec{x}, \tau ) \frac{1}{|x-y|} j^{v}_{ 0 \alpha }( \vec{y}, \tau )
\end{equation}
     Namely, the vortex loop density- density interaction in SS stays  as $ 1/r $ which is the
     same as that in NS ! Therefore, a single vortex loop energy
     and the critical transition temperature $ T_{3dxy} $ in Fig.1
     is solely determined by the superfluid density $ \rho_{s} $
     independent of any other parameters in Eqn.\ref{is},
     except that the vortex core of the vortex loop is much larger
     than that in a superfluid \cite{qglprl}.

      In a cylindrical geometry used in the torsional oscillator
      experiment, the vortex loops will become straight vortex line
      alone the rotational axis. In the SF phase, a single vortex
      costs a lot of energy $ E^{SF}_{v} = \frac{ \rho^{SF}_{s} h^{2} }{
      4 \pi m^{2} } ln \frac{R}{\xi_{SF}} $ where $ m $ is the mass of He
      atom, $ R $ is the system size and $ \xi_{SF} \sim a  $ is the core size of
      the vortex. This energy determines the critical velocity in SF
      $ v^{SF}_{c} > 30cm/s $.
      Because the long distance behavior of SS is more or less the same as
      SF, we can estimate its single vortex energy $ E^{SS}_{v} = \frac{ \rho^{SS}_{s} h^{2} }{
      4 \pi m^{2} } ln \frac{R}{ \xi_{SS} } $ where $ \rho^{SS}_{s} $ is the  {\em global} superfluid density inside
      the SS. We expect the core size of a supersolid vortex
       $ \xi_{SS} \sim 1/\Lambda  \gg 1/k_{r} \sim a  \sim  \xi_{SF} $. So inside the  SS vortex core,
      we should also see the lattice structure of $ n $ \cite{core}. This is similar to the
      phenomenon that DW ordered  states were detected
      in the vortex core of high temperature
      superconductors \cite{high,dmrg}. In fact, because $ \psi(x) $ stands for vacancies or interstitials,
      we expect that $ \xi_{SS} $ should be of the order of
      the average spacing between the interstitials or vacancies in the SS.
      It is interesting to see if neutron or light scattering  experiments can test this prediction.
      Compared to $ E^{SF}_{v} $, there are two reductions, one is the superfluid density,
      another is the increase of the vortex core size $ \xi_{SS} \gg \xi_{SF} $. These two factors contribute
      to the very low critical velocity $ v^{SS}_{c} \sim 30 \mu m/s
      $. Of course, the reduction from the increase of the vortex
      core is negligible because of the logarithmic dependence.

     Integrating out the $ a_{ \alpha \beta } $,
     we get the vortex loop current- current interaction:
\begin{equation}
    2 \pi^{2} j^{v}_{\alpha \beta} (-\vec{q},-\omega_{n} )
    D_{\alpha \beta, \gamma \delta} ( \vec{q}, \omega_{n} )  j^{v}_{\gamma \delta }(\vec{q},\omega_{n} )
\end{equation}
     where $ D_{\alpha \beta, \gamma \delta} ( \vec{q}, \omega_{n} )
     =( \delta_{\alpha \gamma} \delta_{\beta \delta}- \delta_{\beta \gamma} \delta_{\alpha \delta}
     - \frac{ q_{\beta} q_{\delta} }{q^{2}} \delta_{\alpha \gamma}
     - \frac{ q_{\alpha} q_{\gamma} }{q^{2}} \delta_{\beta \delta}
     + \frac{ q_{\alpha} q_{\delta} }{q^{2}} \delta_{\alpha \delta}
     + \frac{ q_{\beta} q_{\gamma} }{q^{2}} \delta_{\alpha \delta }
     )  D_{t}(\vec{q}, \omega_{n} ) $ where  $ D_{t}(\vec{q}, \omega_{n}
     ) $ is the $ a_{t} $ propagator. Defining
      $  \Delta D_{t}(\vec{q}, \omega_{n}) = D^{SS}_{t}(\vec{q}, \omega_{n})-D^{SF}_{t}(\vec{q}, \omega_{n}) $
      as the difference between the $a_{t} $ propagator in the SS and the SF,
      then from Eqn.\ref{prop}, we can get:
\begin{equation}
  \Delta D_{t} = \frac{ a^{2} \rho^{2}_{s} q^{4} }{ \kappa \rho_{n}
  ( \omega^{2}_{n} + v^{2}_{+} q^{2} ) ( \omega^{2}_{n} + v^{2}_{-} q^{2} )( \omega^{2}_{n} + v^{2}_{s} q^{2} ) }
\end{equation}

     For simplicity, we just give the expression for the equal time
\begin{widetext}
\begin{equation}
  \Delta D_{t}( \vec{x}-\vec{x}^{\prime},\tau=0 )
   = \frac{ a^{2} \rho^{2}_{s} }{ 4 \pi^{2} \kappa^{2} \rho^{2}_{n} } \frac{
   v_{+} + v_{-} +v_{s} }{ ( v_{+} + v_{-} )( v_{s}+ v_{+})(
   v_{s}+v_{-}) v_{+} v_{-} v_{s}  } \frac{1}{ (  \vec{x}-\vec{x}^{\prime}
   )^{2} }
\end{equation}
\end{widetext}
   Namely, the vortex current-current interaction in SS is stronger
   than that in the SF with the same parameters $ \kappa, \rho_{s} $ !

\section{ X-ray scattering from the SS }

    Let's look at the prediction of our theory on X-ray
    scattering from the SS. For a lattice with $ j=1, \cdots, n $ basis located
    at $ \vec{d}_{j} $, the geometrical structure factor at the reciprocal lattice vector $  \vec{K} $
    is $ S ( \vec{K} )= \sum^{n}_{j=1} f_{j}( \vec{K} ) e^{i \vec{K} \cdot \vec{d}_{j}}  $ where $ f_{j}
    $ is the atomic structure factor of the basis at $ \vec{d}_{j}
    $.  The X-ray scattering amplitude $ I( \vec{K}) \sim | S ( \vec{K} ) |^{2} $.
    Again, we discuss SS-v and SS-i respectively.

\subsection{ X-ray scattering from the SS-v }
    In this case, because the superfluid density wave simply sits on
    the $ n $ lattice, so the X-ray scattering is very similar to
    that from NS at mean field level. However, as shown in C,
    quantum  and thermal fluctuations will still make the X-ray scattering from the SS-v
    different from that from the NS.

\subsection{ X-ray scattering from the SS-i }
    In this case, as shown in section VI, the superfluid density wave
    is shifted from $ n $ lattice, so the X-ray scattering is different from
    that from NS even at mean field level.
    For simplicity, we first take the $ sc $ lattice to explain the main points, then list the X-ray
    scattering from all the other lattices classified in section VI.

    (a) Simple Cubic lattice.   For the SS in the $ sc $ lattice,
    as shown in (c2) of the last section, the local superfluid density attains its
    maximum at the dual lattice points of the $ sc $ lattice. Then
    $ \vec{d}_{1}=0, \vec{l}_{1}= \frac{a}{2} ( \vec{i} + \vec{j} + \vec{k} ),
      \vec{K} = \frac{2 \pi}{ a } ( n_{1} \vec{i} + n_{2} \vec{j} + n_{3}  \vec{k} ) $,
    then taking the ratio of the geometric structure factor of SS over that of the NS
    $ S_{SS} ( \vec{K})/S_{NS} =1+ f (-1)^{ n_{1} + n_{2} + n_{3} } $
    where $ f \sim \rho^{max}_{s} \sim  a^{2} $. It is $ 1+ f $ for even $
    \vec{K} $ and $ 1 - f $ for odd $ \vec{K} $.

    (b) Triangular lattice.  $ \vec{d}_{1}=0, \vec{l}_{1}= \frac{1}{3} ( \vec{a}_{1} + \vec{a}_{2}
    ),  \vec{l}_{2}= \frac{2}{3} ( \vec{a}_{1} + \vec{a}_{2} ), \vec{K}
     =  n_{1} \vec{b}_{1} + n_{2} \vec{b}_{2} $,
    then taking the ratio of the geometric structure factor of SS over that of the NS
    $ S_{SS} ( \vec{K})/S_{NS} =1+ 2 f \cos \frac{ 2 \pi}{3} ( n_{1} + n_{2}
    ) $. This result could be relevant to possible 2d excitonic
    superfluid in electron-hole bilayer system to be briefly
    mentioned in section X.

    (c) $ hcp $ lattice.  $ hcp $ lattice is not a Bravais lattice.
    In NS,
    $ \vec{d}_{1}=0, \vec{d}_{2}= \frac{1}{3} ( \vec{a}_{1} + \vec{a}_{2}
    )+ \vec{a}_{3}/2, \vec{K}
     =  n_{1} \vec{b}_{1} + n_{2} \vec{b}_{2} + n_{3} \vec{b}_{3}$,
     then  $ S_{NS} ( \vec{K}) = 1+ e^{i 2 \pi ( \frac{ n_{1}+n_{2}}{3} + \frac{n_{3}}{2} ) } $.
     In the SS,  there are 2 more additional basis at
     $   \vec{l}_{1}= \frac{2}{3} ( \vec{a}_{1} + \vec{a}_{2}
    )+ \vec{a}_{3}/4, \vec{l}_{2}= 3 \vec{a}_{3}/4 $, then
     $ S_{SS} ( \vec{K}) = S_{NS}( \vec{K} ) + f e^{i 2 \pi ( \frac{
     2( n_{1}+n_{2})}{3} + \frac{n_{3}}{4} )} + f  e^{i \frac{ 3 \pi}{2} n_{3} }
     =  S_{NS}( \vec{K} ) + f  e^{-i  \frac{
     \pi}{2} n_{3} } S^{*}_{NS}( \vec{K} ) $.

    (d) $ bcc $ lattice. We think $ bcc $ lattice as a $ sc $
    lattice plus a basis, $ \vec{d}_{1}=0, \vec{d}_{2}= \frac{a}{2} ( \vec{i} + \vec{j} + \vec{k} ),
      \vec{K} = \frac{2 \pi}{ a } ( n_{1} \vec{i} + n_{2} \vec{j} + n_{3}  \vec{k} )
      $, Then $ S_{NS} ( \vec{K} ) =1+ (-1)^{ n_{1} + n_{2} + n_{3} } $
     which is $ 2 $ for even $ \vec{K} $ and $ 0 $ for odd $ \vec{K} $.
     In the SS, there are 3 more additional basis at $  \vec{l}_{1}= \frac{a}{2} ( \vec{i} +
     \vec{j} ), \vec{l}_{2}= \frac{a}{2} ( \vec{i} + \vec{k} ),  \vec{l}_{3}= \frac{a}{2} ( \vec{j} + \vec{k}
     ) $, then $ S_{SS} ( \vec{K} ) = S_{NS} ( \vec{K} )+ f [ (-1)^{ n_{1} + n_{2} } +  (-1)^{ n_{1} + n_{3} } +
     (-1)^{ n_{2} + n_{3}} ] $.

    (e) $ fcc $ lattice. We think $ fcc $ lattice as a $ sc $
    lattice plus 4 basis located  at $ \vec{d}_{1}=0, \vec{d}_{2}= \frac{a}{2} ( \vec{i} +
     \vec{j} ), \vec{d}_{3}= \frac{a}{2} ( \vec{i} + \vec{k} ),  \vec{d}_{4}= \frac{a}{2} ( \vec{j} + \vec{k}
     ), \vec{K} = \frac{2 \pi}{ a } ( n_{1} \vec{i} + n_{2} \vec{j} + n_{3}  \vec{k} ) $,
      then $ S_{NS} ( \vec{K} ) =1+ [ (-1)^{ n_{1} + n_{2} } +  (-1)^{ n_{1} + n_{3} } + (-1)^{ n_{2} + n_{3}
     } ] $. In the SS, there is one more additional basis located  at $
      \vec{l}_{1}= \frac{a}{2} ( \vec{i} + \vec{j} + \vec{k} ) $, then
     $ S_{SS} ( \vec{K})= S_{NS} ( \vec{K} )  + f (-1)^{ n_{1} + n_{2} + n_{3}} $. It is easy to see that
     in $ bcc $ and $ fcc $ lattices, we need simply exchange $ d $
     vectors for the NS and the $ l $ vectors for SFDW.

    We conclude that the elastic X-ray
    scattering intensity from the SS-i has an additional modulation
    over that of the NS. The modulation amplitude is proportional
    to the maxima of the superfluid density $ \rho^{max}_{s} \sim  a^{2} $ which is
    the same as the NCRI observed in the PSU's torsional oscillator experiments.

\subsection{  Debye-Waller factor in the X-ray scattering from the SS-v and SS-i }

    It is known that due to zero-point quantum motion in any NS at very low temperature, the X-ray
    scattering amplitude $ I (\vec{G}) $ will be diminished by a Debye-Waller (DW) factor
    $ \sim e^{- \frac{1}{3} G^{2} \langle u^{2}_{\alpha}\rangle } $ where $ u_{\alpha} $
    is the lattice phonon modes in Eqn.\ref{ss1}. In Eqn.\ref{ss1},
    if the coupling between the $ \vec{u} $ and $ \theta $ were
    absent, then the DW factor in the SS would be the same as
    that in the NS.
    By taking the ratio $ I_{SS}( \vec{G})/I_{NS}( \vec{G}) $ at a given reciprocal lattice vector
    $ \vec{G} $, then this DW factor drops
    out. However, due to this coupling,  the $ \langle u^{2}_{\alpha}\rangle
    $ in SS is different than that in NS, so the DW factor
    will {\sl not} drop out in the ratio.
    In this subsection, we will calculate this ratio and see how to take care of this factor
    when comparing with the X-ray scattering data.

  As identified below Eqn.\ref{orderhe},  the density order parameter at the reciprocal lattice vector $ \vec{G} $ is
  $ \rho_{ \vec{G} }( \vec{x},\tau )  = e^{ i \vec{G} \cdot \vec{u}( \vec{x},\tau )  } $,
  then $ \langle \rho_{ \vec{G} }( \vec{x},\tau ) \rangle = e^{-\frac{1}{2}
  G_{i}G_{j} \langle u_{i} u_{j} \rangle } $. The Debye-Waller factor:
\begin{equation}
   I( \vec{G} ) =
  | \langle \rho_{ \vec{G} }( \vec{x},\tau ) \rangle |^{2}=  e^{- G_{i}G_{j} \langle u_{i}( \vec{x},\tau ) u_{j}( \vec{x},\tau ) \rangle }
\label{dw}
\end{equation}
  where the phonon-phonon  correlation  function is:
\begin{equation}
   \langle u_{i} u_{j} \rangle =  \langle u_{l}u_{l} \rangle  \hat{q}_{i}\hat{q}_{j} +  \langle u_{t}u_{t} \rangle ( \delta_{ij}- \hat{q}_{i}\hat{q}_{j} )
\label{uu}
\end{equation}
   where $ \hat{q}_{i}\hat{q}_{j}= \frac{ q_{i} q_{j} }{ q^{2}} $.  .

   Then substituting Eqn.\ref{uu} into Eqn.\ref{dw} leads to:
\begin{equation}
  \alpha( \vec{G} ) = I_{SS}( \vec{G} )/I_{NS}(
  \vec{G})= e^{-\frac{1}{3} G^{2}[ \langle u^{2}_{l}( \vec{x},\tau ) \rangle_{SS}- \langle u^{2}_{l}( \vec{x},\tau ) \rangle_{NS} ] }
\end{equation}
  where the transverse mode drops out,  because it stays the same in the SS and in the NS.

  Defining $ ( \Delta u^{2})_{l}( \vec{q}, i \omega_{n}  )=
  \langle | u_{l}( \vec{q}, i \omega_{n}  )|^{2} \rangle_{SS}- \langle | u_{l}( \vec{q}, i \omega_{n}  )|^{2} \rangle_{NS} $,
  $ ( \Delta u^{2} )_{l}( \vec{q} )= \sum_{ i \omega_{n} } ( \Delta u^{2})_{l}(
     \vec{q}, i \omega_{n} ) $  and
  $  ( \Delta u^{2})_{l}= \langle u^{2}_{l}( \vec{x},\tau ) \rangle_{SS}- \langle u^{2}_{l}( \vec{x},\tau ) \rangle_{NS}
  =  \int \frac{ d^{3}q }{ (2 \pi)^{d} } \frac{1}{\beta} \sum_{ i \omega_{n} } ( \Delta u^{2})_{l}(
     \vec{q}, i \omega_{n})  = \int \frac{ d^{3}q }{ (2 \pi)^{d} }( \Delta u^{2} )_{l}( \vec{q} )   $,  it is easy to see:
\begin{widetext}
\begin{equation}
  ( \Delta u^{2})_{l} =  \int \frac{ d^{3}q }{ (2
  \pi)^{3} } \frac{1}{\beta} \sum_{ i \omega_{n} } \frac{- a^{2}
  q^{2} \omega^{2}_{n} }{ [ ( \kappa \omega^{2}_{n} + \rho_{s} q^{2} ) (  \rho_{n} \omega^{2}_{n} + ( \lambda+2 \mu
    ) q^{2} ) + a^{2} q^{2} \omega^{2}_{n} ][  \rho_{n} \omega^{2}_{n} + ( \lambda+2 \mu
    ) q^{2} ]}
\label{long}
\end{equation}
\end{widetext}
    Obviously, $ ( \Delta u^{2})_{l} < 0   $, namely,
    the longitudinal vibration amplitude in SS is {\em smaller} that that
    in NS. The $ \alpha( \vec{G} )( T=0 )= e^{-\frac{1}{3} G^{2}( \Delta
    u^{2})_{l}}  > 1 $.
    This is expected, because the SS state is the ground
    state at $ T < T_{SS} $, so the longitudinal vibration amplitude
    should be reduced compared to the corresponding NS with the same
    parameters $ \rho_{n}, \lambda, \mu $.

    After evaluating the frequency summation in Eqn.\ref{long}, we get:
\begin{eqnarray}
  ( \Delta u^{2})_{l}(T)  =  \int \frac{ d^{3}q }{ (2
  \pi)^{3} } \frac{1}{ \rho_{n} }[ \frac{ \coth \beta v_{+} q/2 }{ 2
  v_{+} q } - \frac{ \coth \beta v_{p} q/2 }{ 2 v_{p} q }  \nonumber  \\
  -  (\frac{ v^{2}_{s}-v^{2}_{-} }{ v^{2}_{+}-v^{2}_{-} })(
  \frac{ \coth \beta v_{+} q/2 }{ 2
  v_{+} q } - \frac{ \coth \beta v_{-} q/2 }{ 2 v_{-} q } )]
\label{finite}
\end{eqnarray}

   At $ T=0 $, the above equation simplifies to:
\begin{eqnarray}
  ( \Delta u^{2})_{l}( T=0)  =  \int \frac{ d^{3}q }{ (2
  \pi)^{3} } \frac{1}{ \rho_{n} }[ \frac{ 1 }{ 2
  v_{+} q } - \frac{ 1 }{ 2 v_{p} q }  \nonumber  \\
  -  ( \frac{ v^{2}_{s}-v^{2}_{-} }{ v^{2}_{+}-v^{2}_{-} })(
  \frac{ 1 }{ 2 v_{+} q } - \frac{1 }{ 2 v_{-} q } )]  ~~~~~~~~~~~~~~~~~  \nonumber \\
  = - \frac{ ( v_{+}+v_{-}-v_{p}-v_{s}) }{ ( v_{+}+v_{-} ) v_{p} }
  \frac{ \Lambda^{2} }{ 8 \pi^{2} \rho_{n} } ~~~~~~~~~~~~~~~~ \nonumber \\
   =  - \frac{ a^{2}}{ \kappa \rho_{n} }  \frac{1}{
  (v_{+}+v_{-}+v_{p}+v_{s})( v_{+}+v_{-} ) v_{p} } \frac{ \Lambda^{2} }{ 8 \pi^{2} \rho_{n}
  } < 0
\label{zero}
\end{eqnarray}
   where  $ \Lambda \sim 1/a $ is the ultra-violet cutoff and
   we have used the fact $ v_{+}+v_{-} > v_{p} + v_{s} $.

  By subtracting Eqn.\ref{zero} from Eqn.\ref{finite}, we get
\begin{eqnarray}
  ( \Delta u^{2})_{l}( T )-( \Delta u^{2})_{l}( T=0)  = ~~~~~~~~~~~~~~~~~~~~~~~~~~ \nonumber  \\
  \frac{ ( v_{+}-v_{p}
  )( v_{+} + v_{p} )- ( v_{s}-v_{-})(v_{+}+v_{-} )}{ (v_{+}+v_{-})
  v_{+} v_{-} v^{2}_{p} }  \frac{ ( k_{B} T )^{2} }{ 12
  \rho_{n} } > 0
\end{eqnarray}
    Namely, the difference in the ratio  will {\em decrease } as $ T^{2} $ as the
    temperature increases. Of course, when $ T $ approaches $ T_{SS} $ from below, the
    difference vanishes, the $ \alpha( \vec{G} ) $ will approach $ 1 $ from above, the SS turns into a NS.


\subsection{ Density-density correlations }

    The density-density correlation function in the SS is:
\begin{equation}
    \langle \rho_{ \vec{G} }( \vec{x}, t ) \rho^{*}_{ \vec{G} }( \vec{x}^{\prime}, t^{\prime} )
    \rangle= e^{-\frac{1}{2}  G_{i}G_{j}
    \langle ( u_{i}(\vec{x}, t )-u_{i}(\vec{x}^{\prime}, t^{\prime} ))
      ( u_{j}(\vec{x}, t )-u_{j}(\vec{x}^{\prime}, t^{\prime} )) \rangle }
\end{equation}
   where $ t $ is the real time.

    For simplicity, we only evaluate the equal-time correlator
    $  \langle \rho_{ \vec{G} }( \vec{x}, t ) \rho^{*}_{ \vec{G} }( \vec{x}^{\prime}, t
    ) \rangle = \langle \rho_{ \vec{G} }( \vec{x}, \tau ) \rho^{*}_{ \vec{G} }( \vec{x}^{\prime},
    \tau ) \rangle  $ where $ \tau $ is the imaginary time. It is instructive to compare the density order in
    SS with that in a NS by looking at
    the ratio of the density correlation function in the SS over the NS:
\begin{equation}
     \alpha_{\rho}( \vec{x}- \vec{x}^{\prime} )= \langle \rho_{ \vec{G} } \rho^{*}_{ \vec{G} }\rangle_{SS}/ \langle \rho_{ \vec{G} } \rho^{*}_{ \vec{G} }\rangle_{NS}
     =  e^{-\frac{1}{6} G^{2} \Delta D_{\rho} ( \vec{x}- \vec{x}^{\prime} ) }
\label{ratiorho}
\end{equation}
      It is easy to find that
\begin{equation}
   \Delta D_{\rho} ( \vec{x}- \vec{x}^{\prime} ) = \int \frac{ d^{3}q }{(2
   \pi)^{3} } ( 2-e^{i \vec{q} \cdot (\vec{x}-\vec{x}^{\prime})}-
                 e^{-i \vec{q} \cdot (\vec{x}-\vec{x}^{\prime})}) ( \Delta u^{2})_{l}( \vec{q} )
\end{equation}
     where $ ( \Delta u^{2})_{l}( \vec{q} ) $ is defined above Eqn.\ref{long} and is the integrand in
     Eqn.\ref{finite}.

     At $ T=0 $, the above equation can be simplified to
\begin{equation}
     \Delta D_{\rho}( \vec{x}- \vec{x}^{\prime} )
     = \frac{ ( v_{+} + v_{-} -v_{p}-v_{s} )}{ (v_{+}+v_{-}) v_{p} }
     \frac{1}{ 2 \pi^{2} \rho_{n} } \frac{1}{( \vec{x}- \vec{x}^{\prime} )^{2} }
\end{equation}

     So we conclude that $  \alpha_{\rho}( \vec{x}- \vec{x}^{\prime} ) < 1 $, namely, the density order in SS is {\em weaker }
     than the NS with the corresponding parameters $ \rho_{n}, \lambda, \mu $.
     This is expected because the density order in the SS is
     weakened by the presence of moving vacancies.

    Unfortunately, so far, the X-ray scattering data is limited to
    high temperature $ T> 0.8 K > T_{SS} $ \cite{simmons}. X-ray scattering experiments on lower
    temperature $ T< T_{SS} $ are being performed to test these predictions.

\section{ Specific heat in the SS }

     It is well known that  at low $ T $, the specific heat in the NS is $
     C^{NS}= C^{NS}_{lp} + C^{NS}_{tp} + C_{van} $
     where $ C^{NS}_{lp} =
     \frac{ 2 \pi^{2}}{15} k_{B} ( \frac{k_{B}T}{\hbar v_{lp} })^{3}
     $  is from the longitudinal phonon mode and $ C^{NS}_{tp} =
     2 \times \frac{ 2 \pi^{2}}{15} k_{B} ( \frac{k_{B}T}{\hbar v_{tp} })^{3} $
     is from the two transverse phonon modes, while $ C_{van} $ is
     from the vacancy contribution. $ C_{van} $ was calculated in
     \cite{heatwave} by assuming 3 different kinds of models for the
     vacancies. So far, there is no consistency between the calculated $ C_{van}
     $ and the experimentally measured one \cite{heatwave,ander}.
     The specific heat in the SF $ C^{SF}_{v}= \frac{ 2 \pi^{2}}{15} k_{B} ( \frac{k_{B}T}{\hbar v_{s} })^{3} $
     is due to the SF mode $ \theta $. In this subsection, we focus on
     the specific heat inside the SS. From Eqn.\ref{is}, we can find the specific heat in the SS:
\begin{equation}
 C^{SS}_{v}= \frac{ 2 \pi^{2}}{15} k_{B} ( \frac{k_{B}T}{\hbar
v_{+} })^{3} + \frac{ 2 \pi^{2}}{15} k_{B} ( \frac{k_{B}T}{\hbar
v_{-} })^{3} + C^{tp}
 \label{spec}
\end{equation}
    where $ C^{tp} $ stands for the contributions from teh
    transverse phonons which are the same as those in the NS ( Fig.7 ).

     It was argued in \cite{qglprl}, the critical regime of finite temperature  NS to
     SS transition in Fig.1 is much narrower than the that of SF to
     the NL transition, so there should be a jump in the specific
     heat at $ T=T_SS $ ( Fig.7 ). Eqn.\ref{spec} shows that at $ T < T_{SS}
     $, the specific heat still takes $  \sim T^{3} $ behavior  and is
     dominated by the $ \omega_{-} $ mode in Fig.6.

\begin{figure}
\includegraphics[width=6cm]{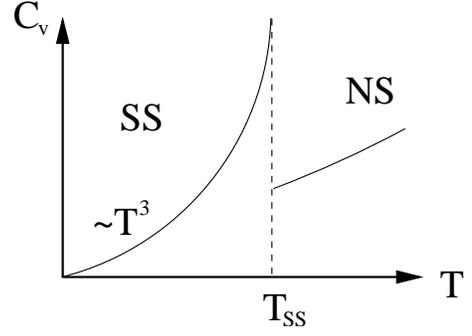}
\caption{ The specific heat of a supersolid } \label{fig7}
\end{figure}

     From Eqn.\ref{spec}, it is easy to evaluate the entropy inside the SS:
\begin{equation}
 S_{SS}(T_{SS})= \frac{ 2 \pi^{2}}{45} k_{B} ( \frac{k_{B} T_{SS}
 }{\hbar})^{3}( \frac{1}{ v^{3}_{+} }+ \frac{1}{ v^{3}_{-} } ) + S^{tp}
 \label{enss1}
\end{equation}
      where $ S^{tp} $ is the entropy due to the 2 transverse modes.

      The {\em excess} entropy  due to the vacancy condensation is:
\begin{equation}
 \Delta S = \int^{T_{SS}}_{0} dT  C_{van}/T = \frac{ 2 \pi^{2}}{45} k_{B} ( \frac{k_{B} T_{SS}
 }{\hbar})^{3}( \frac{1}{ v^{3}_{+} }+ \frac{1}{ v^{3}_{-} }- \frac{1}{ v^{3}_{lp} })
 \label{enss2}
\end{equation}

       Obviously,  $ \Delta S >0 $ is due to the lower branch $
       v_{-} < v_{lp} $. At $ a=0 $, the above equation reduces to $
       \Delta S= \frac{ 2 \pi^{2}}{45} k_{B} ( \frac{k_{B} T_{SS} }{\hbar v_{s}
       })^{3} $ which is simply the vacancy condensation into SF.
       Using the molar volume $ v_{0} \sim 20 cm^3/mole $ of solid $^{4} He $ and $ T_{ss} \sim 100 mK $, we can estimate
       the $ \Delta S $ per mole is $ \sim 10^{-5} R $ where $ R $
       is th gas constant. This estimate is 3 orders magnitude
       smaller than that in \cite{entropy} where the SS state was
       taken simply as the boson condensation of non-interacting
       vacancies. Our estimate is indeed consistent with recent
       experiment on specific heat \cite{heat}.
     The linear $ \sim T $ behaviour found in heat capacity
     experiment  can only be due to disorder or $ ^{3} He $
     impurities.


\section{ Elementary excitations in Fulde - Ferrell-Larkin-Ovchinnikov (FFLO) state of superconductors  }

  The QGL theories constructed for the supersolid in the previous sections may also be used to study the elementary excitations in the inhomogeneous
  Fulde - Ferrell-Larkin-Ovchinnikov (FFLO) state of superconductors \cite{ff,lo,ganfflo,fflorev}.
  As argued in the following, the FFLO state can be taken as a fermionic supersolid.
  When the number of spin up electron is equal to the number of down spin electron $ n_{\uparrow}=n_{\downarrow} $, the pairing is at $ \vec{k}=0 $ only.
  If there is a mismatch $ \delta n= n_{\uparrow} -n_{\downarrow} $, then pairing will shift to a non-zero momentum $ q_0=k_{F \uparrow}- k_{F \downarrow} $.
  By using the GL theory near the normal to the FFLO state ( Fig.8 ),
  at a mean field level, the authors in  \cite{fflo} found the most favorable lattice structures of the FFLO state
  is the stripe state (LO state) in large number of parameter regimes. The FFLO state maybe considered as a weak coupling ( or fermionic ) analog
  of the ( bosonic ) supersolid. Indeed, the Fig.8 is similar to the Fig.2 after identifying the chemical potential difference $ \delta \mu $
  with the pressure $ p $.
  It is important to point out a few important differences
  between the GL for the bosons developed in this paper and the GL for the FFLO state used in \cite{fflo}: (1) In the former, due to the lack of the $ Z_2 $ symmetry, there is a cubic term in Eqn.\ref{sl}. But in the latter, due to the presence of the $ Z_2 $ symmetry, there is no such a cubic term.
  (2) In the latter, the transition from the normal state to the FFLO state is a classical Lifshitz type transition, so there is no zero momentum pairing
  as shown in \cite{fflo} and in Eqn.\ref{orderfflo}, in contrast to the second equation in Eqn.\ref{orderhe}  where there is always a zero momentum
  BEC condensation. (3) In the latter, there are large number of unpaired fermions which are the normal component of the system. Their distributions also
  take the same lattice structure as the FFLO state. They play a similar role as the normal lattice component in the first equation
  in \ref{orderhe}, but they form extended Bloch waves in the underlying FFLO lattice and are integrated out in the GL theory in \cite{fflo}.

\begin{figure}
\includegraphics[width=2in]{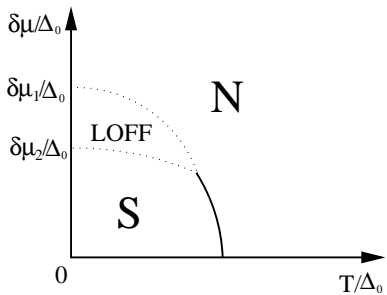}
\caption{The phase diagram of FFLO state. $ \delta \mu $ is the
difference of the two chemical potentials of two species of fermions, $ T $ is the temperature, $ \Delta_{0} $ is the
energy gap at the balanced case $ \delta \mu=0 $. }
\end{figure}

  So far all the analysis in a FFLO state\cite{ff,lo,ganfflo,fflorev,fflo} are only at a mean field level.
  In this appendix, we study the elementary excitations above the mean field solutions.
  Just from symmetry breaking point of views, the FFLO state breaks both the translational symmetry and the $ U(1) $ symmetry, so it
  also has both the diagonal order and off-diagonal order, so the low energy modes should contain both the lattice phonon mode and the
  superfluid phonon mode. The pairing order parameter can be written as:
\begin{eqnarray}
 \psi_{FFLO} ( \vec{x}, \tau )  =  \psi_{0}  \sum^{\prime}_{\vec{G}}
      e^{i \vec{G} \cdot
     ( \vec{x} + \vec{u}( \vec{x}, \tau ) )}
\label{orderfflo}
\end{eqnarray}
     where $ G=q_0 $, the $ \psi_{0}( \vec{x}, \tau )= \Delta e^{ i \theta(  \vec{x}, \tau ) }  $
     is the pairing order parameter, $ \vec{u}( \vec{x}, \tau )  $ are the lattice phonon modes. Note the absence of the zero momentum pairing.
     For a charged system such as a electron system, due to the Higgs mechanism, the Goldstone mode $ \theta(  \vec{x}, \tau ) $ will be just eaten by the gauge field.
     However, for a neutral system such as  the pairing between two species of fermions with unequal
     populations in  cold atom systems across a Feshbach resonance \cite{son},
     the Goldstone mode $ \theta(  \vec{x}, \tau ) $ stays.
 The coupling between the two phonon modes in Eqn.\ref{orderfflo} are also described by a equation similar to Eqn.\ref{is}. The only difference is
 that the lattice structure is a LO state instead of a isotropic solid. After taking this difference into account, the elementary excitations
 inside the FF state  are similar to those in the Fig.6a and the corresponding spectral weights can be worked out similarly. The experimental signatures of the
 elementary excitations can also be worked out similarly.

  There exist other kinds of fermionic supersolids in addition to the FFLO state.
  In some fermionic systems, it is easy to see the coexistence of CDW and
  Superconductivity ( SC), so " fermionic supersolid " phases are common.
  For instance,  a quasi-two-dimensional system
  NbSe2 has a transition to an incommensurate CDW phase at some high temperature  $ T_{CDW} $ and then a
  transition to a phase with coexisting CDW and SC order at a lower
  temperature $ T_{SC} $. The  CDW is a pairing in particle-hole channel at
  $ 2k_{F} $, its order parameter is $ \psi_{CDW}= \langle
  c^{\dagger}_{\sigma}(\vec{k}) c_{\sigma}( \vec{k}+\vec{Q} ) \rangle $ where $
  \vec{Q } $ is the ordering wave vector of the CDW.
  The SC is a pairing in particle-particle channel also across the Fermi surface.
  Its order parameter is $ \psi_{SC}= \langle c^{\dagger}_{\uparrow}(\vec{k}) c^{\dagger}_{\downarrow}( - \vec{k}
  ) \rangle $. Both order parameters are composite order parameters. Different
  parts of Fermi surface can do the two jobs separately ( see, for
  example, \cite{neto} ).
  In contrast to the bosonic SS where there is a density operator $ n $ and a complex order parameter $ \psi $,
  both order parameters  $ \psi_{CDW} $ and $ \psi_{SC} $  are complex order
  parameters.  We can see that
  although the formation of a supersolid in a bosonic system has completely different mechanism than in fermionic case, from symmetric point of view,
  the QGL theories constructed developed in previous sections could also be used to
  describe the interplay between the two complex order parameters, the
  properties of the "fermionic supersolid" (FSS) and the transition from
  the $ FSS $ to the CDW. However, the fermionic excitations near the nodes
  maybe also be important  and need to be taken into account\cite{neto}.

\section{ Conclusions}


    In this paper, assuming there are quantum fluctuations generated vacancies or interstitial at zero temperature \cite{dislocation},
    we constructed a two component QGL theory to map out a possible global phase diagram, analyze carefully the conditions
    for the existence of the supersolid  and study all the phases and phase transitions in a unified view.
    The only new parameters introduced in the GL theory in this paper is the
    coupling  $ g $ and $ v $ between the $ n $ sector ( or normal solid part ) and the $ \psi $ sector ( or the
    superfluid part ) in Eqn.\ref{int}. We investigated the SS state from both the SF and the NS side
    and found completely consistent description of the properties of
    the SS state.  Starting from the SF side with increasing the pressure, we developed the theory basing on the two
    facts (1) there is a roton minimum in
    the superfluid state (2) the instability to solid formation is
    driven by the gap diminishing at the roton minimum.   By increasing the pressure from
    the superfluid side, there are two possible scenarios (1) the SF to the C-NS transition in Fig.1 was
    described by the QGL action Eqn.\ref{sfdensity}  first derived in this
    paper (2) the SF to the SS  transition in Fig.2 is
    a simultaneous  combination of the
     SFDW  transition in the $ \psi $ sector driven by
     the roton condensation  at $ k_{0}=k_{r} $ and the NS transition in the
    $ n $ sector driven by the divergence in  the structure function $ k_{0}= k_{n}=k_{r} $.
    The superfluid becomes
    a SS at lower temperature and a NS at higher
    temperature ( Fig.2 ). Then we also approached the SS state from the NS
    side. Depending on the sign and strength of the coupling $ g $
    between the solid and superfluid, we also found two possible scenarios:
    (1) If $ |g| $ is  sufficiently small ( Fig.4 ), then
    the resulting solid at $ T=0, p_{c1} < p < p_{c2} $ is a commensurate normal solid ( C-NS ).
    The SS state does not exist as a ground state. However, it may still exist as a metastable
    state. The QGL action to describe this SF to NS transition in Fig.1 was developed in Sec.II.
    (2)  If $ |g| $ is  sufficiently large ( Fig.4 ), the resulting solid at $ T=0,  p_{c1} < p < p_{c2}  $ is
    an incommensurate solid  with zero point quantum fluctuations generated vacancies if
    it is negative and interstitials if it is positive ( Fig.4 ).
    The condensation of the vacancies or  interstitials lead to the formation of the
    SS-v and SS-i respectively. The SS state has lower energy than the NS state at $ T=0 $.
    The $ T_{SS-v} $ ( $ T_{SS-i} $ ) is an effective
    measure of the strength of the
    interaction $ g $ in the SS-v ( SS-i ) supersolid. There is no
    particle-hole symmetry relating $ T_{SS-v} $ to  $ T_{SS-i} $.
    Many physical consequences came out of this single parameter $ g $.
    Our results on supersolid should be independent of many microscopic details
    and universal.

    Just like the SF is a uniform two-component phase consisting of superfluid and
    normal component at any finite temperature, the SS state
    is a {\em uniform } two-component phase consisting of a superfluid density wave (SFDW) and
    a normal solid component even at zero temperature.
    The SFDW in the SS-v coincides with the underlying normal solid.
    While the SFDW in the SS-i
    state is just a dual lattice to the underlying normal solid.
    This important fact leads to one of the key predictions in this paper: the X-ray
    scattering intensity from the SS-v is similar to that of NS at
    mean field level, while the X-ray
    scattering intensity from the SS-i ought to have an additional modulation
    over that of the NS. The modulation amplitude is proportional
    to the Non-Classical Rotational-Inertial (NCRI) observed in the
    torsional oscillator experiments.  However, the X-ray scatterings from SS-v and SS-i
    will be modified by the Debye-Waller factor calculated in sec. IV.
    The NS-v ( NS-i ) to SS-v (SS-i )
    transition is described by a 3d $ XY $ model with much narrower critical regime (Fig.4).

     We also studied the zero temperature quantum
     phase transition from the SS to the NS driven by the pressure
     near the upper critical pressure $ p=p_{c2} $ in Fig.5. We found
     that the coupling to the quantum fluctuation of the underlying
     lattice is irrelevant, so the transition stays the same
     universality class as the superfluid to Mott insulator
     transition in a 3 dimensional rigid optical lattice. The finite
     temperature transition from the SS to the NS in Fig.1 was
     studied previously in \cite{dor} and in \cite{elas} in
     different contexts. It was found that the coupling to classical
     elastic degree of freedoms will not change the
     universality class of the 3D $ XY $ transition. However, we
     found that the coupling to quantum lattice phonons is very
     important inside the SS  and leads to two "supersolidon" modes $
     \omega_{\pm}=v_{\pm} q $ ( one upper branch and one lower branch ) shown in Fig.2.
     Their corresponding spectral weights are also worked out. The transverse modes
     in the SS stays the same as those in the NS. Detecting the two
     supersolidon modes with the corresponding spectral weights, especially, the lower branch $ \omega_{-} $
     mode by neutron scattering \cite{neutron} or acoustic wave attention
     experiments \cite{acou} is a smoking gun experiment to prove or disprove
     the existence of the SS in helium $ ^{4} He $. The $ \omega_{-}
     $ is estimated to considerably  lower than the sound speed
     in the superfluid. Then we calculated the experimental
     signature of the two supersolidon modes. We found that the longitudinal
     vibration in the SS is smaller than that in the NS ( with the same corresponding solid parameters ),
     so the Debye-Waller  factor at a given reciprocal lattice vector is larger than
     that in the NS. The density-density correlation function in the
     SS is weaker than that in the NS. By going the to the dual
     vortex loop representation, we found the vortex loop
     density-density interaction in SS stays the same as that in the
     SF ( with the same corresponding superfluid parameters ), so
     the vortex loop energy and the corresponding SS to NS transition
     temperature is solely determined by the superfluid density and
     independent of any other parameters. The vortex current-current
     interaction is stronger than that in the SF.
     The specific heat   in the SS is still given by the sum from the transverse phonons and the two  supersolidon
     modes and still shows $ \sim  T^{3} $ behaviors.
     The supersolidon part is dominated by the lower branch.
     The NCRI is only weakly anisotropic in the SS phase for $ hcp $ lattice.
     In principle, all these predictions can be  tested by experimental techniques such as X-ray scattering,
     neutron scattering, acoustic wave attenuations \cite{acou,heatwave} and heat capacity.

      It may be instructive to make some analogy of Fig.1 and   Fig.2 at $ T < T_{SF} $ to Type-I
    and type-II superconductors with the pressure  $ p $ playing the role of the magnetic field $ H $:
    Fig.1 is similar to Type-I superconductor with SF identified as the Messiner state, the NS
    as the normal state, the critical pressure $ p_{c} $ identified as the critical magnetic field $
    H_{c} $. Fig.2 is similar to Type II superconductor with SF identified as the Messiner state, the SS as the mixed
    vortex lattice state which also breaks both translational order and the global $ U(1) $ symmetry, the NS
    as the normal state, the lower and upper critical pressures $ p_{c1} $ and $ p_{c2} $
    identified as the  lower and upper critical magnetic fields $ H_{c1} $ and $ H_{c2} $.
    In superconductors, it is the $ \kappa=\lambda/\xi $ to
    determine Type I and Type II and if the vortex lattice is a
    stable intermediate state or not as the magnetic field is increased. In Helium 4, it is the sign
    and strength of the coupling constant $ g $ in Eqn.\ref{int} to
    determine the Fig.1 and Fig.2 and if the SS is a
    stable intermediate state or not as the pressure is increased.
    { \em So the pressure $ p $ and the coupling $ g $ in the formation of SS-v play the role of the magnetic
    field $ H $ and $ \kappa $ in the formation of the mixed state of superconductors. }
    Note that in superconductors, $ H $ and $ \kappa $ are
    two independent parameters, in $ ^{4} He $, $ p $ and $ g $ are
    also two independent parameters.

   The GL theory developed  in this paper put the competing orders of superfluid and solid in
   the unified framework.  We suggest that even supersolid may
   not be realized in $ ^{4}He $ system, it has its own intrinsic,
   scientific interests  and may be realized in other continuous bosonic and fermionic
   systems. For example, in symmetric electron-hole bilayer systems \cite{short1,short2,long}, it was
   shown in \cite{ess} that it is quite possible that there may a narrow window of
   ESS where both order parameters are non-vanishing $ \langle \psi\rangle \neq 0, \langle n_{\vec{G} } \rangle \neq0 $
   intervening between the ESF and two weakly coupled Winger crystal.
   A similar GL theory can be used to study the elementary excitations inside a FFLO state and some
   fermionic  systems with coexistence of CDW and superconductivity.

{\bf Acknowledgement}

    I thank  P. W. Anderson, M. Chan, T. Clark, Milton Cole, B.
    Halperin, Jason Ho,  T. Leggett, T. Lubensky, Mike
    Ma, G. D. Mahan, S, Sachdev and F. C. Zhang for helpful
    discussions, A. T. Dorsey for pointing out Ref. \cite{roton2} to
    me.  I also thank the hospitality of Y. Chen, Z. Wang
    and F. C. Zhang during my visit at Hong Kong University,
    Yu Lu during my visit at Institute for Theoretical Physics in Beijing, China.
    The research at KITP was supported by the NSF grant No.
    PHY99-07949.  The research at KITP-C is supported by
    the Project of Knowledge Innovation Program (PKIP) of Chinese Academy of Sciences.

\appendix

\section{ Discussions on a tight-binding toy Supersolid ground state wavefunction  }

 The Ginzburg-Landau theory constructed in the main text is based on
 order parameters and symmetries. It should hold irrespective many
 microscopic details such as what is the mechanism responsible for
 the formation of the supersolid.
 Despite there are many microscopic calculations for $ ^{4} He $, constructing
 a microscopic theory for supersolid is very difficult. In this
 appendix, I will discuss a well known toy SS wavefunction and clarify a few concepts
 related to {\em global} phase-number uncertainty relation and the role of
 vacancies or interstitials in the formation of SS. We also clarify the physical
 meaning of the order parameters $ n $ in Eqn.\ref{sl} and $ \psi $ in Eqn. \ref{sfnormal}.
 However, because the toy wavefunction may
 miss some important physics in bulk $^{4}He$ systems. For example,
 due to the very peculiar potential well in the solid $^{4} He $ which
 has a local shallow maximum at the lattice site, the tight binding
 model maybe crude, so the discussion is  intuitive and instructive.
 Eqn.\ref{ss} is also a ground state wavefunction, so does not include the lattice phonon
 interaction.

 The toy  wavefunction of a supersolid takes the BCS like form
\begin{equation}
    |SS\rangle = \prod^{N}_{i=1} ( u + v b^{\dagger}_{i} ) |0\rangle
\label{ss}
\end{equation}
    where $ N $ is the number of sites,
    $ u \neq 0 $ and $ |u|^{2}+ |v|^{2}=1 $. If setting $ u=0 $, the state reduces to
    a commensurate solid without any vacancies $ |CS\rangle= \prod^{N}_{i=1} b^{\dagger}_{i}
    |0\rangle $. The commensurate solid (CS) is an exact eigenstate of the boson number operator
    $ N_{b}=\sum^{N}_{i=1} n_{i} $ with  the eigenvalue $ N_{b}=N $, so has no chance to become phase ordered.
    Adding a superfluid component to the CS leads to the SS in
    Eqn.\ref{ss}. If setting $ v=0 $, then the state reduces to the vacuum state $
    |0\rangle $.

    If setting $ u=\cos \frac{\theta}{2}, v=\sin \frac{\theta}{2}
    e^{i \phi} $, then Eqn.\ref{ss} becomes:
\begin{equation}
    |SS, \phi\rangle = \prod^{N}_{i=1} ( \cos \frac{\theta}{2} + \sin \frac{\theta}{2}
    e^{i \phi}  b^{\dagger}_{i} ) |0\rangle
\label{ssp}
\end{equation}
    where $ \theta \neq \pi $.

    By construction, the state has the translational order with the average boson
    density $ \langle n_{i} \rangle = |v|^{2}=\sin^{2} \theta/2 $, so the average vacancy density
    is $ |u|^{2}=\cos^{2} \theta/2 $. The total number of bosons $ N_b= N \sin^{2} \theta/2 $,
    the number of vacancies $ N_v= N \cos^{2} \theta/2 $.
    It is easy to see that $ |SS\rangle $ also has the ODLRO with
    $ \langle b_{i} \rangle=u^{*}v= \frac{1}{2} \sin \theta e^{i \phi} $, so $ |SS\rangle $ is indeed a supersolid state.
    The angle $ \theta $  controls the magnitude, while the phase $
    \phi $ controls the phase of the condensation. Defining $ b_{k=0} = \frac{1}{\sqrt{N}} \sum^{N}_{i=1} b_{i}
    $ which satisfy the boson commutation relation $ [ b_{0},
    b^{\dagger}_{0}]=1 $, the boson operator at zero momentum is $ n_{0}= b^{\dagger}_{k=0}b_{k=0} $,
    the total number of
    bosons at the zero momentum state is $ N_{0}= \langle SS|n_{0}|SS\rangle= \frac{N}{4}
    \sin^{2} \theta= N_{b} \cos^{2} \theta/2 < N_{b}= N \sin^{2} \theta/2 < N $.
    Similarly $ N_{0}= \langle SS|n_{0}|SS\rangle= \frac{N}{4}
    \sin^{2} \theta= N_{v} \sin^{2} \theta/2 < N_{v}= N \cos^{2} \theta/2 < N $.
    So the condensate is less than both the boson density and the vacancy density.
    At integer filling $ n=1 $, the non-interacting BEC state $ |SF\rangle = \frac{1}{ \sqrt{ N ! }} (
    b^{\dagger}_{k=0} )^{N} |0\rangle $, then $ N_{0}=
    \langle SF|n_{0}|SF\rangle=N_{b}=N $.
    Obviously, this non-interacting BEC state is not included in the family in the
    Eqn.\ref{ss}.

    A supersolid state $ |SS, N_{b} \rangle $ with $ N_{b} $ bosons is given by:
\begin{equation}
    |SS, N_{b} \rangle= \int^{2 \pi}_{0} \frac{ d \phi}{2
     \pi} e^{-i N_{b} \phi} |SS, \phi\rangle
\label{ssn}
\end{equation}
     where the total boson number $ N_{b} $ and the global phase $ \phi $ are two Hermitian conjugate
     variables satisfying the commutation relation: $ [N_{b}, \phi]=i \hbar  $.
     It leads to the uncertainty relation $ \Delta N_{b} \Delta \phi \geq 1 $.

     It is easy to see $ \langle N_{b}\rangle = N |v|^{2}=N \sin^{2} \theta/2,
     \Delta N_{b} = \sqrt { \langle N^{2}_{b} \rangle-\langle N_{b}\rangle^{2} } =
     \sqrt{N} |u v| = \sqrt{N} \frac{1}{2} |\sin \theta |,  \Delta N_{b} /\langle N_{b}\rangle  = \frac{1}{\sqrt{N}} |u/v|=
     \frac{1}{\sqrt{N}} |\cot \theta/2 | $.
     If $ \theta \neq \pi $, the absolute boson number fluctuation
     $ \Delta N_{b} \sim \sqrt{N} $ is quite large, so $ \Delta \theta $
     could be quite small, so one can get a phase ordered state. On
     the other hand, the relative boson number fluctuation
     $  \Delta N_{b} /\langle N_{b}\rangle  \sim \frac{1}{\sqrt{N}}
     $ is quite small, so one can still measure the average boson
     number accurately. The first quantization form of the Eqn.\ref{ssn} can be derived by the same method
     used in \cite{wave}.

     Because in the SS state, there is a global phase ordering in $
     \phi $, so its conjugate variable is the total number of
     particles $ N_{b} $ as shown in Eqn.\ref{ssn}.
     The local tunneling
     or exchanging processes  stressed in \cite{leg} may not cause the total number fluctuations,
     therefore may not cause the global phase ordering leading to the
     supersolid phase. Eqn.\ref{ss} implies that the vacancies in an incommensurate solid
     could lead to the formation of a supersolid. If this indeed
     happens, in the GL theory constructed for SS-v in the main text, the lattice sites are represented by $ n(x) $,
     while the vacancies are represented by $ \psi $. Of course, the trial ground wavefunction \ref{ss} does not
     include fluctuations, so no information on the lattice phonons and superfluid phonons, therefore no supersolidons in the Fig.6a.
     A toy wavefunction for SS-i was not written
     down so far, because the interstitials are moving between lattice
     sites, so there is no tight binding limit. The interstitial
     case is described by the extended Bloch state in Section IV-B.
     In the SS-i, the bosons are represented by $ n(x) $, while the interstitials are represented by $ \psi $.

\section{ Comparisons with Supersolids on lattices }

     The extended boson Hubbard model ( EBHM ) with various kinds of interactions,
     on all kinds of lattices and at different filling factors is described by the following
     Hamiltonian \cite{boson}:
\begin{eqnarray}
   H  & = & -t \sum_{ \langle ij \rangle } ( b^{\dagger}_{i} b_{j} + h.c. )
          - \mu \sum_{i} n_{i} + \frac{U}{2} \sum_{i} n_{i} ( n_{i} -1 )
                                  \nonumber   \\
      & + & V_{1} \sum_{ \langle ij\rangle } n_{i} n_{j}  + V_{2} \sum_{ \langle\langle ik\rangle\rangle } n_{i} n_{k} + \cdots
\label{boson}
\end{eqnarray}
    where $ n_{i} = b^{\dagger}_{i} b_{i} $ is the boson density, $ t $ is the nearest neighbor hopping amplitude.
    $ U, V_{1}, V_{2} $ are onsite, nearest neighbor (nn) and next nearest neighbor (nnn) interactions respectively,
    the $ \cdots $ may include further neighbor interactions and possible ring-exchange interactions.
     A supersolid in Eqn.\ref{boson} is defined as the
     simultaneous orderings of ferromagnet in the $ XY $ component ( namely, $ \langle b_{i} \rangle \neq 0 $ )
     and CDW in the $ Z $ component.
    In \cite{substrate}, we studied all the possible phases and phase transitions in
    the EBHM in  bipartite lattices such as a honeycomb and square  lattice near half
    filling. We show that there are two consecutive transitions at zero temperature
    {\em driven by the chemical potential}: in the Ising limit, a
    Commensurate-Charge Density Wave (CDW) at half filling to
    a narrow window of CDW supersolid, then to
    an Incommensurate-CDW ; in the easy-plane limit, a
    Commensurate-Valence Bond Solid (VBS) at half filling to
    a narrow window of VBS supersolid, then to an Incommensurate-VBS.
    The first transition is second order  in the same universality class as the Mott to insulator transition
    , therefore has the exact critical exponents $
    z=2, \nu=1/2, \eta=0 $ with logarithmic corrections, while the second one is first order. Liu
    and Fisher \cite{liu} also studied the C-CDW to the CDW-SS transition
    and concluded that $ z=1 $ in contrast to $ z=2 $ achieved in
    \cite{substrate}.
    We found that the phase diagram in the
    Ising limit is similar to the reentrant "superfluid"  in a narrow region of coverages in the
    second layer of $^{4}He $
    adsorbed on graphite detected by Crowell and Reppy's torsional oscillator experiment in 1993 \cite{he}.
    Indeed, the data in the torsional oscillator experiment in \cite{he}
    do not show the characteristic form for a 2d $^{4} He $ superfluid film,
    instead it resembles that in \cite{chan} characteristic of a supersolid in terms of the gradual onset temperature of the
    NCRI, the unusual temperature dependence of $ T_{SS} $ on the coverage.
    Of course, both experiments may due to phase separations instead
    of the SS phase. very recently, the author studied various kinds
    od supersolids in frustrated lattices such as triangular and
    kagome lattices \cite{frus}.

     By comparing  Eqns.\ref{sl}, \ref{sfnormal},\ref{int} in the continuum with the
     Eqn.\ref{boson} on the lattice, one can see the crucial difference between a lattice system with a
     continuous system:  Eqn \ref{sl} stands for a spontaneously
     formed lattice with phonon excitations. While Eqn.\ref{boson}
     is a fixed external lattice with no such phonon excitations.
     This crucial difference is responsible for the difference \cite{bragg} in the elementary excitation spectra in
     the two kinds of supersolid shown in Fig.6.
     So the reentrant lattice SS discussed in
     \cite{liu,substrate,frus} is different from the bulk  $ ^{4}He $ SS state discussed in this
     paper, although both kinds of supersolids share some interesting common properties.
     In both cases, the SF to SS transition is driven by the closing of the gap of
     the roton minimum, so the transition could be first or second order at mean field level.
     However, due to the fluctuation of the underlying superfluid order, the transition turns out to be
     the first order as confirmed by the QMC simulations.
     On the lattice, the manifold of the roton
     minimum consists of discrete points due to the lattice symmetry. However, in $^{4} He $, as shown in sections III and IV,
     the manifold of the roton
     minimum is a continuous surface, so the transition must be first order.
     order. In the former, there is a periodic substrate or spacer potential
      which breaks translational symmetries at the very beginning.
      The filling factor is controlled by an external chemical
      potential.  At integer filling factors  or half filling factor, there are particle-hole symmetry (PH) for excitations which
      ensures the number of particles is equal to  that of vacancies and adding interstitials is equivalent to adding
      vacancies respectively \cite{substrate,frus}.
      While in the latter, the lattice results from a spontaneous
      translational symmetry breaking driven by the pressure as shown in the Fig.2,
      if there are vacancies or interstitials in the ground state has to be self-determined by ground state energy
      minimization.
      There is usually no particle-hole symmetry for
      excitations. Due to this
      absence of symmetry, the number of vacancies is usually not
      equal to that of interstitials. So the theory developed in
      this paper on bulk $ ^{4}He $ is  different from the lattice theory developed in \cite{liu,substrate}.
      As shown in the appendix A and in \cite{substrate,frus},
      one common fact of both supersolids is that both are due to vacancies or interstitials.
      Lattice supersolids can also be described by doping the adjacent CDW either by vacancies or interstitials, so
      are classified as two types $ SS-v $ and $ SS-i $. In the
      hard-core limit, $ SS-v $ and $ SS-i $ are simply related by
      P-H transformation. However, as shown in section IV, in
      Helium-4 supersolid, there is no particle-hole symmetry relating $ T_{SS-v} $ to  $ T_{SS-i} $ ( Fig.3 ) !
      {\em So the coupling constant $ g $ in Helium 4 plays a similar
      role as the chemical potential $ \mu $ in the lattice models, the gap $ \Delta(p) $
      in the NS-PH which tunes the distance from the NS-PH to the SF plays a similar role as
      the gap in the CDW which tunes the distance from the CDW to the superfluid }.
      It was shown in \cite{frus}, the lattice
      supersolids existing at commensurate 1/2 filling factors in frustrated lattices such
      as triangular lattice is just the coexistence of SS-v and SS-i.
      Combined with the results in \cite{substrate},
      we conclude that $ ^{4}He $ supersolid can exist both in bulk and on substrate,
      while although $ H_2 $ supersolid may not exist in the bulk,
      but it may exist on wisely chosen substrates.
      Lattice supersolid could also be realized in optical lattices in ultra-cold
      atomic experiments \cite{bragg}. However, in both continuum
      and on lattices, SS states could be unstable against phase
      separations. For example, the vacancies or interstitials in the
      in-commensurate solid can simply move to the boundary of the
      sample instead of boson condensation, namely, it will turn
      into a commensurate solid. This case is included in the C-NS case in the paper anyway.
      Due to its negative compressibility, the instability of lattice SS  against phase
      separation was demonstrated in some lattice models in \cite{hardsoft}.



\begin{thebibliography}{99}
\bibitem{ring} Feynman originally conceived the roton as drifting
   vortex loop. But this point of view is very controversial. If taking this view, then the roton
   condensation can be considered as
   vortex loop condensation.
\bibitem{core} The core energy is always much smaller than the
  superflow energy even for large size of vortex core.
\bibitem{tonyp} In a general direction which is neither along $ \hat{z} $ nor in $ xy $ plane,
    exactly perpendicular or parallel conditions are not met.  However, one can still use
    the notions of quasi-transverse and quasi-longitudinal modes
    as done in the Fig.10 in the first reference in \cite{sound}.
    One can project the direction onto the z-axis and xy-plane,
    the elastic constants in the velocity expressiones are just some effective average of
    the 5 different elastic constants.
\bibitem{string} In fact, if setting $ a=0 $, Eqn.\ref{dual} can be
written in a Lorentz invaiant form: $ {\cal L}_{v}= \frac{1}{3
K_{\mu} } H^{2}_{\mu \nu \lambda} + i 2 \pi a_{\mu \nu} j^{v}_{\mu
\nu} $ where $ H_{\mu \nu \lambda}=
 \partial_{\mu} a_{\nu \lambda}+
\partial_{\nu} a_{ \lambda \mu} + \partial_{\lambda} a_{\mu \nu} $
is  gauge invariant under the gauge transformation $a_{\mu \nu}
\rightarrow  a_{\mu \nu} + \partial_{\mu} \chi_{\nu}- \partial_{\nu}
    \chi_{\mu} $. However, this neat form is not of practical use in
    the presence of the $ a $ coupling term in Eqn.\ref{dual}.
\end{thebibliography}
\end{document}